\documentclass[fleqn,usenatbib]{mnras}
\usepackage{newtxtext,newtxmath}

\usepackage[T1]{fontenc}

\DeclareRobustCommand{\VAN}[3]{#2}
\let\VANthebibliography\thebibliography
\def\thebibliography{\DeclareRobustCommand{\VAN}[3]{##3}\VANthebibliography}
\defcitealias{2022MNRAS.510.1480K}{Paper~I}

\usepackage{graphicx}	
\usepackage{amsmath}	
\usepackage{threeparttable}
\usepackage{xcolor}
\usepackage{caption}

\title[Evolution of the quasar 0858$-$279]{Parsec-scale evolution of the gigahertz-peaked spectrum quasar PKS\,0858$-$279}

\author[Kosogorov et al.]{\parbox{\textwidth}{
N. A. Kosogorov,$^{1,2,3}$\thanks{E-mail: nakosogorov@gmail.com}
Y. Y. Kovalev,$^{4,2,1}$
M. Perucho,$^{5,6}$
Yu. A. Kovalev$^{2}$
}
\vspace{0.4cm}\\
\parbox{\textwidth}{
$^{1}$Moscow Institute of Physics and Technology, Institutsky per.~9, Dolgoprudny 141700, Russia\\
$^{2}$Lebedev Physical Institute of the Russian Academy of Sciences,
Leninsky prospekt 53, 119991 Moscow, Russia\\
$^{3}$Cahill Center for Astronomy and Astrophysics, MC 249-17 California Institute of Technology, Pasadena, CA 91125, USA\\
$^{4}$Max-Planck-Institut für Radioastronomie, Auf dem H\"ugel 69, D-53121 Bonn, Germany\\
$^{5}$Departament d'Astronomia i Astrof\'{\i}sica, Universitat de Val\`encia, C/ Dr. Moliner, 50, E-46100, Burjassot, Val\`encia, Spain\\
$^{6}$Observatori Astron\`omic, Universitat de Val\`encia, C/ Catedr\`atic Jos\'e Beltr\'an 2, E-46980, Paterna, Val\`encia, Spain
}}

\date{Accepted 2024 January 03. Received 2023 December 28; in original form 2023 September 29}

\pubyear{2023}

\begin{document}
\label{firstpage}
\pagerange{\pageref{firstpage}--\pageref{lastpage}}
\maketitle

\begin{abstract}

We conducted multi-epoch, multi-frequency parsec-scale studies on the gigahertz-peaked spectrum quasar PKS\,0858$-$279 with the Very Long Baseline Array (VLBA). Our observations on 2005-11-26 elucidated a weak core, characterized by an inverted spectrum, and a distinctly bent jet that exhibited a notable bright feature in its Stokes I emission. Through comprehensive analysis of polarization and spectral data, we inferred the formation of a shock wave within this feature, stemming from interactions with a dense cloud in the ambient medium. In this paper, VLBI-\textit{Gaia} astrometry further reinforces the core identification. With a deep analysis of six additional VLBA epochs spanning from 2007 to 2018, we observed that while the quasar's parsec-scale structure remained largely consistent, there were discernible flux density changes. These variations strongly imply the recurrent ejection of plasma into the jet. Complementing our VLBA data, RATAN-600 observations of the integrated spectra suggested an interaction between standing and travelling shock waves in 2005. Moreover, our multi-epoch polarization analysis revealed a drastic drop in rotation measure values from 6000~rad/m$^2$ to 1000~rad/m$^2$ within a single year, attributable to diminishing magnetic fields and particle density in an external cloud. This change is likely instigated by a shock in the cloud, triggered by the cloud's interaction with the jet, subsequently prompting its expansion. Notably, we also observed a significant change in the magnetic field direction of the jet, from being perpendicular post its observed bend to being perpendicular prior to the bend~--- an alteration possibly induced by the dynamics of shock waves.

\end{abstract}



\section{Introduction}

\begin{figure*}
	\includegraphics[width=\linewidth,trim={0cm 0cm 0cm 0cm},clip]{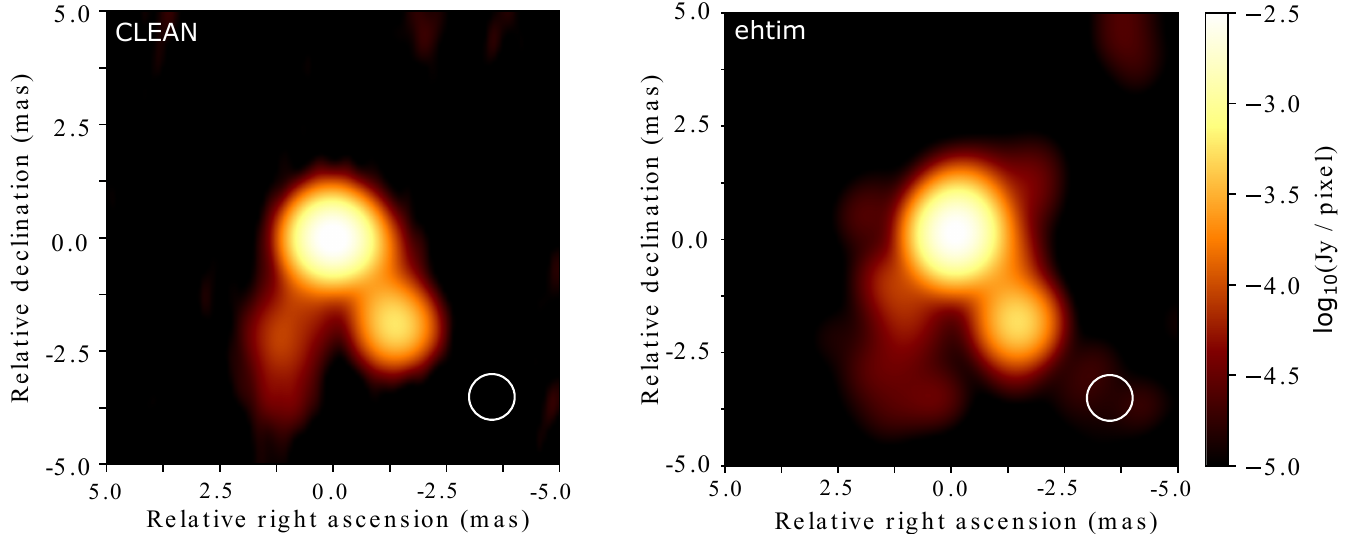}
	\caption{High-resolution images of the source at 22.2 GHz from the 2005-11-26 epoch of the VLBA observation obtained with different reconstruction algorithms. The image obtained with the CLEAN algorithm is presented on the left. It was reconstructed with a circular Gaussian beam. The image reconstructed with the eht-imaging package is shown on the right. 
    It was blurred with a circular Gaussian beam taken from the image on the left. The beams are shown as white circles in the bottom right corner at the FWHM level. The FWHM axis size for both beams is 1.0 mas.}
    \label{fig:eht_im}
\end{figure*}

The evolution of active galactic nuclei (AGN) is pivotal to understanding the formation of galaxies and studying the environment in which they evolve. Gigahertz-peaked sources (GPS) are key candidates believed to be in early developmental phases. Probing these sources can deepen our comprehension of galaxies during their initial stages of growth. While the general framework of AGN evolution encompasses vast cosmic timescales, our specific focus in this article is on a timeframe of roughly 20 years. This allows us to highlight short-term dynamics that may offer insights into broader evolutionary patterns. Several studies have cast light upon GPS sources, unravelling their temporal properties \citep[e.g.][]{1998PASP..110..493O,1998A&A...329..845C, 2003ASPC..300...71L, 2008ApJ...680..911S,2021A&ARv..29....3O}. These sources manifest an inverted spectrum below the turnover frequency, attributed to synchrotron self-absorption or free-free absorption. By scrutinizing the short-term variations in their properties and spectra, we aim to address many challenges surrounding GPS sources, such as classification issues, their interaction dynamics with the surrounding medium, and the inherent nature of absorption mechanisms.

Very Long Baseline Interferometry (VLBI) offers an unparalleled perspective, allowing us to delve deep into the processes occurring near the central engines of quasars. VLBI emphasizes explosive events and the intricate birth and formation dynamics of jets. These jets can interact dynamically with their surrounding environment. The capacity of VLBI to distinguish moving jet components offers a detailed glimpse into the kinematics of relativistically moving plasma \citep[e.g.][]{2009AJ....138.1874L}. Typically bright knots, whose origin remains an open question, are commonly observed. In some cases, they have superluminal apparent speeds far from the core, but they can also be quasi-stationary \citep[e.g.][]{2011MmSAI..82...33L, 2016A&A...592A..22H}. Several models have been proposed to explain the properties of the moving plasma \citep[e.g.][]{1979ApJ...232...34B, 1985ApJ...298..114M}. Furthermore, numerical simulations are used to investigate how jets evolve and to exhibit these bright features \citep[e.g.][]{1999MNRAS.308.1069K,
1997ApJ...482L..33G, 
2001ApJ...549L.183A,
2016A&A...588A.101F,2021NewAR..9201610K}. 

In a previous article \citep{2022MNRAS.510.1480K}, which we will refer to as \citetalias[][]{2022MNRAS.510.1480K}, we investigated the properties of quasar PKS\,0858$-$279 in detail. The source captured our interest due to its high amplitude variability over timescales of several months from the RATAN-600 data, while displaying extended structure in the early VLBI images. We resolved both the core and a prominent bright jet feature. Analyzing the spectral and polarization properties, we found out that a shock wave was formed in that feature. This shock formation may explain the notably stronger magnetic field in the jet feature, estimated to be $\sim$1G, compared to the core's magnetic field of about 0.1~G. Furthermore, from the multi-frequency light curves, we discerned a variability Doppler factor of around 2. This estimation method leverages the Doppler boosting of flux density variations in the quasar’s jet, representing one approach to find the Doppler boosting factor (see eq. (3) in \citetalias[][]{2022MNRAS.510.1480K}). In this work, our aim is to understand the changes in the properties of the source over time, pinpoint the source of variability, ascertain the kinematics of the jet plasma, and track the evolution of the brightest jet feature. To achieve these goals, we examine the source evolution using multi-epoch polarization VLBI observations.

This paper is structured as follows: in \autoref{sec:data}, we provide the observation details and the data reduction process.  \autoref{gaia} showcases the joint analysis of optical \textit{Gaia} positional observations compared with the VLBI ones. In \autoref{sec:res}, we present the results of the VLBA and RATAN-600 data analysis, including Stokes I images, modelling of the source structure, kinematics, spectral properties, variability and polarization properties. In \autoref{sec:discus}, we discuss our results in light of possible scenarios of the changes which took place during the time-span of our observations. Finally, \autoref{sec:summary} sums up all the findings. Through the paper, the model of flat $\mathrm{\Lambda C D M}$ cosmology is used with $H_0 = 68\,\mathrm{km\,s}^{-1}$, $\Omega_M$ = 0.3 and $\Omega_\Lambda=0.7$~\citep{2016A&A...594A..13P}. The position angles are measured from north through east.

\section{Observations and Data reduction}
\label{sec:data}

\subsection{VLBA observations}
\label{sec:vlba_obs}

\begin{table*}
	\centering
	\caption{Summary of the VLBA observations for 0858$-$279.}
	\label{tab:obs}
	\begin{tabular}{llll}
		\hline\hline
		Epoch & Frequency (GHz) & Data source & Comments on the data \\ 
	    \hline
		2005-11-26 & 1.5--22 & NRAO archive BK128 & bad weather at BR\\   
        2007-02-09 & 1.5--22 & NRAO archive BK133A & communication at MK, receiver at LA \\   
		2007-05-23 & 1.5--22 & NRAO archive BK133B & data absence at OV, power glitches at KP,\\  
		& & & bad weather at LA and NL, receiver at PT \\
		2007-08-03 & 1.5--22 & NRAO archive BK133C & data absence at BR \\   
		2017-05-27 & 8.3 & Astrogeo database and NRAO archive UF001I & possible issues with amplitude calibration \\   
		2018-04-08 & 8.3 & Astrogeo database and NRAO archive UG002E &  \\   
		2018-07-21 & 8.3 & Astrogeo database and NRAO archive BP222A & \\   
		\hline
	\end{tabular}
	\begin{tablenotes}
            \item \emph{Note.} The table shows the corresponding epoch of the VLBI observations, frequency range, source where the data can be found and problems with the data, which led to flagging. We made use of the publicly available data from the NRAO archive with the project codes given in the table and from the Astrogeo VLBI FITS image database (see details in \autoref{sec:vlba_obs}). For entries with project codes starting with ``BK'' from the NRAO archive, we utilized correlated data. In contrast, the Astrogeo database entries reference imaged data, although the same observations can also be found in the NRAO archive with the project codes listed in the table.
        \end{tablenotes}
\end{table*}

As part of a dedicated campaign to study the quasar, it was observed with the Very Long Baseline Array (VLBA) of the National Radio Astronomy Observatory (NRAO) on 26~November~2005, 9~February~2007, 23~May~2007 and 3~August~2007. The project codes are BK128, BK133a, BK133b and BK133c. Details of the observation and the reduction of the BK128 project data can be found in \citetalias[][]{2022MNRAS.510.1480K}. Here we describe the BK133 project data. The observations were made in a dual-polarization mode. The observing bands were centred at 1.5~GHz, 2.3~GHz, 4.8~GHz, 8.3~GHz, 15.4~GHz and 22.2~GHz. Each band comprised four 8~MHz-wide frequency channels (IFs) per polarization. We combined all IFs at all frequencies in the consequent analysis, except for the analysis presented in \autoref{sec:pol_prop}. In the polarization analysis, the four lowest frequency bands were split into two sub-bands, two IFs per sub-band, centred at 1.41~GHz, 1.66~GHz, 2.28~GHz, 2.39~GHz, 4.64~GHz, 5.04~GHz, 8.15~GHz, 8.50~GHz. The data at each VLBA station were recorded with 2-bit sampling and a total rate of 256~Mbps in right and left circular polarization. 

The observed calibrators consisted of the
fringe check, cross-polarization and band-pass calibrator 1226+023 (3C\,273) and the leakage-term calibrator 0919$-$260. Apart from that, the electric vector position angle (EVPA) calibrators included 1328+307 (3C\,286) during all three BK133 epochs, 0851$+$202 (OJ\,287) in February and 0923+392 in May and August.

The initial data processing steps were performed in the Astronomical Image Processing System package~\citep{2003ASSL..285..109G} for each sub-band independently, paying special attention to accurate amplitude calibration. They comprised typical VLBI data processing procedures which can be found in the AIPS cookbook. For the first phase self-calibration step, the CALIB task in AIPS was utilized. We used the Difmap software package \citep{1997ASPC..125...77S} to iteratively do phase and amplitude self-calibration as well as obtain CLEAN images. The resulting visibility amplitude calibration accuracy after a~priori and self-calibration steps was estimated to be approximately 10\,\% for 22~GHz and around 5\,\% for frequencies in the 1.4~GHz to 15~GHz range. These estimations agree well with the calibration accuracy described in the VLBA observational status summary\footnote{\url{https://science.nrao.edu/facilities/vlba/docs/manuals}}.
Typical processing of the 8~GHz astrometric VLBI sessions (\autoref{tab:obs}) might result in somewhat lower accuracy of amplitude calibration due to the IFs being wide-spread across the band and shorter on~source observing time.

Apart from using the traditional CLEAN algorithm \citep[][]{1974A&AS...15..417H}, we used regularised maximum likelihood methods, which are implemented in the eht-imaging package \citep[e.g.][]{2018ApJ...857...23C}, to check the fidelity of our images. In \autoref{fig:eht_im}, we present the images at the highest frequency of our observations of 22 GHz in 2005 reconstructed with these two approaches. When reconstructed with the same beam parameters, the results were consistent, with only minor differences noted in the southeast emission. 

We calibrated the EVPA of 0858$-$279 using 1328+307 for 1.5~GHz and 2.2~GHz bands and 0851+202 and 0923+392 for all other frequencies. The calibration procedure was performed in the same manner as described in the previous analysis. We made use of the calibrators absolute EVPA information from the VLA (Very Large Array) Monitoring Program\footnote{\url{http://www.vla.nrao.edu/astro/calib/polar/}} \citep{taylor2000vlba} and the MOJAVE\footnote{\url{https://www.cv.nrao.edu/MOJAVE/}} (Monitoring Of Jets in Active galactic nuclei with VLBA experiments) observations.

Additionally, we accessed VLBI images of the source at 8.3 GHz from three other epochs: 2017-05-27, 2018-04-08, and 2018-07-21, which are publicly available from the Astrogeo VLBI FITS image database\footnote{\url{http://astrogeo.smce.nasa.gov/vlbi_images/}}. The summary of all VLBI observations used in this paper can be found in \autoref{tab:obs}.

\subsection{RATAN-600 observations}
\label{sec:ratan_obs}

Furthermore, we observed the source at 5-7 frequencies from 1 to 22 GHz with the RATAN-600 radio telescope of the Special Astrophysical Observatory \citep[e.g.][]{1993IAPM...35....7P} since 1997, conducting 2-4 observing sessions annually. Within each session, the continuum spectra for the source were typically measured 2-4 times and averaged. We utilised the data from 1997 to the present date, extending the previously used time period of the first 13 years. The details on the observations, calibration, and presented errors can be found in the corresponding part of \citetalias[][]{2022MNRAS.510.1480K}.

We compared the integrated flux densities from the VLBI images with RATAN-600 flux densities for the nearest dates to their VLBI observation. They turned out to be in good agreement with all epochs but 2017-05-27. This specific epoch demonstrated that the flux density is 30$\%$ lower than the corresponding one from RATAN-600. To account for that, we applied a correction in Difmap to make the integrated flux density consistent. We note that, for some datasets within the Astrogeo database, amplitude calibration might not be the primary focus, especially for astrometric/geodetic-type observations with wide-spread IFs. At the same time, every observing RATAN-600 session utilises observations of several calibrators. Given the meticulous calibration procedures associated with RATAN-600 observations, we suggest that it is the VLBI flux density from this specific dataset in the Astrogeo database that required correction.

\subsection{\textit{Gaia} and VLBI astrometry observations}
\label{gaia_obs}

\begin{table*}
	\centering
	\caption{VLBI (RFC version 2023b) and \textit{Gaia} coordinates and their corresponding errors for 0858$-$279. Right ascension coordinate is given in hours, minutes and seconds. Declination is given in degrees, minutes and seconds. }
	\label{tab:gaia}
	\begin{tabular}{ccccc} 
		\hline\hline
		& Right ascension & RA error (mas) & Declination & Dec error (mas)\\ 
	    \hline
		VLBI & 09:00:40.038786 & 0.13 & $-$28:08:20.34539 & 0.17 \\
        \textit{Gaia} & 09:00:40.038693 & 0.06 & $-$28:08:20.34742 & 0.06 \\
        \hline
	\end{tabular}
\end{table*}
This work used the \textit{Gaia} Data Release 3 (\textit{Gaia} DR3) \citep{2016A&A...595A...1G,2023A&A...674A...1G}. This catalogue provides extremely accurate measurements of optical positions of various sources, including AGNs. Similar submilliarcsecond accuracy can be obtained with the VLBI observations. In further analysis, we used the VLBI-based Radio Fundamental Catalog\footnote{\url{http://astrogeo.smce.nasa.gov/sol/rfc/rfc_2023b/}} to cross-identify and study 0858$-$279.

\section{VLBI-\textit{Gaia} astrometry analysis}
\label{gaia}

In most cases we expect \textit{Gaia} positions to be close to the jet base, within the VLBI core. At the same time, the previous analyses of the VLBI-to-\textit{Gaia} offsets showed that there are about 10\,\% of AGN with significant differences between their optical and VLBI coordinates \citep[e.g.][]{2017MNRAS.467L..71P,2017A&A...598L...1K,2019MNRAS.482.3023P,2019ApJ...871..143P}. These offsets can vary and be as large as tens of mas. Among AGN with significant offsets, the majority of sources display \textit{Gaia} positions further downstream along the jet direction compared to the VLBI coordinates. Such behaviour is expected when strong and extended parsec-scale optical features are found in jets. On the other hand, some sources exhibit offsets in the opposite direction. This can be caused by the core-shift effect or the dominant impact of the accretion disk.

\begin{figure*}
	\includegraphics[width=\linewidth]{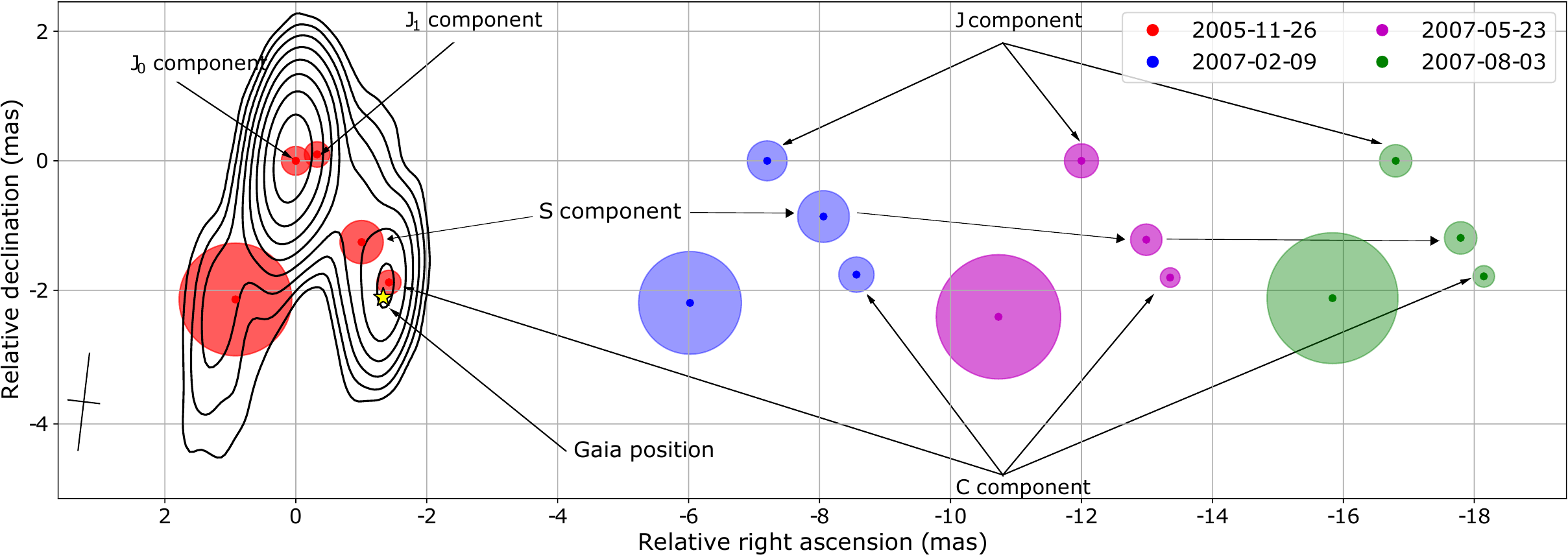}
    \caption{Peak positions and sizes of the Gaussian components obtained at 15~GHz from all the epochs are shown in dots and circles with different colours. The arrows show the dominant jet components and the components associated with the core. We show the full intensity contours for the Stokes $I$ map at 22.2~GHz from the epoch 2005-11-26, as a reference. The contours start at the 2$\sigma_I$ level with the $\times2$ step. The beam is shown in the bottom left corner at the FWHM level. The map at 22~GHz has a peak intensity of 358 mJy/beam, with a map noise level ($\sigma_I$) of 0.8~mJy/beam. The beam parameters are: major axis ($B_\mathrm{maj}$) 1.5~mas, minor axis ($B_\mathrm{min}$) 0.5~mas, and beam position angle ($B_\mathrm{PA}$) $-$6.7~degrees. The components from the three epochs in 2007 are artificially shifted along the right ascension axis. See also \autoref{tab:gauss_comp} for more detail on the parameters of the Gaussian components. We also show the position of the \textit{Gaia} coordinates for the source with a star assuming that the J$_0$ component has the absolute coordinates from the Radio Fundamental Catalog (see details in \autoref{gaia}). One can note that the brightest jet feature is best represented with two components for the epoch 2005-11-26 and one for all other epochs. This can be related to the shock wave evolution in the source (see \autoref{sec:discus}). }
    \label{fig:components}
\end{figure*}

Since 0858$-$279 displays an unusual VLBI parsec-scale radio structure characterized by a weak core and a pronounced jet feature, there was significant interest in comparing its optical coordinates with the VLBI ones. Under the assumption that the RFC VLBI coordinates align with the dominant jet feature J0, the Radio Fundamental Catalog coordinates of the source correspond to this position. This presumption is based on the consistent observation of the feature for over a decade (further elaborated in \autoref{sec:tot_int_struc}). The \textit{Gaia} and VLBI coordinates can be found in \autoref{tab:gaia}. The \textit{Gaia} position does not match the dominant jet feature. The offset between \textit{Gaia} and VLBI is 2.4 $\pm$ 0.2 mas. By understanding this offset and comparing it with the RFC catalog's indication for J0, we positioned the \textit{Gaia} coordinates on our map. As a result, the \textit{Gaia} position appears close to the core region, suggesting its association with the accretion disc. The \textit{Gaia} and the VLBI core positions matched with good accuracy, within the error limits (refer to \autoref{fig:components}). Such a close alignment emphasizes that radiation from the accretion disc and the beggining of the jet dominates at optical wavelengths.

\begin{figure*}
	\includegraphics[width=0.45\linewidth]{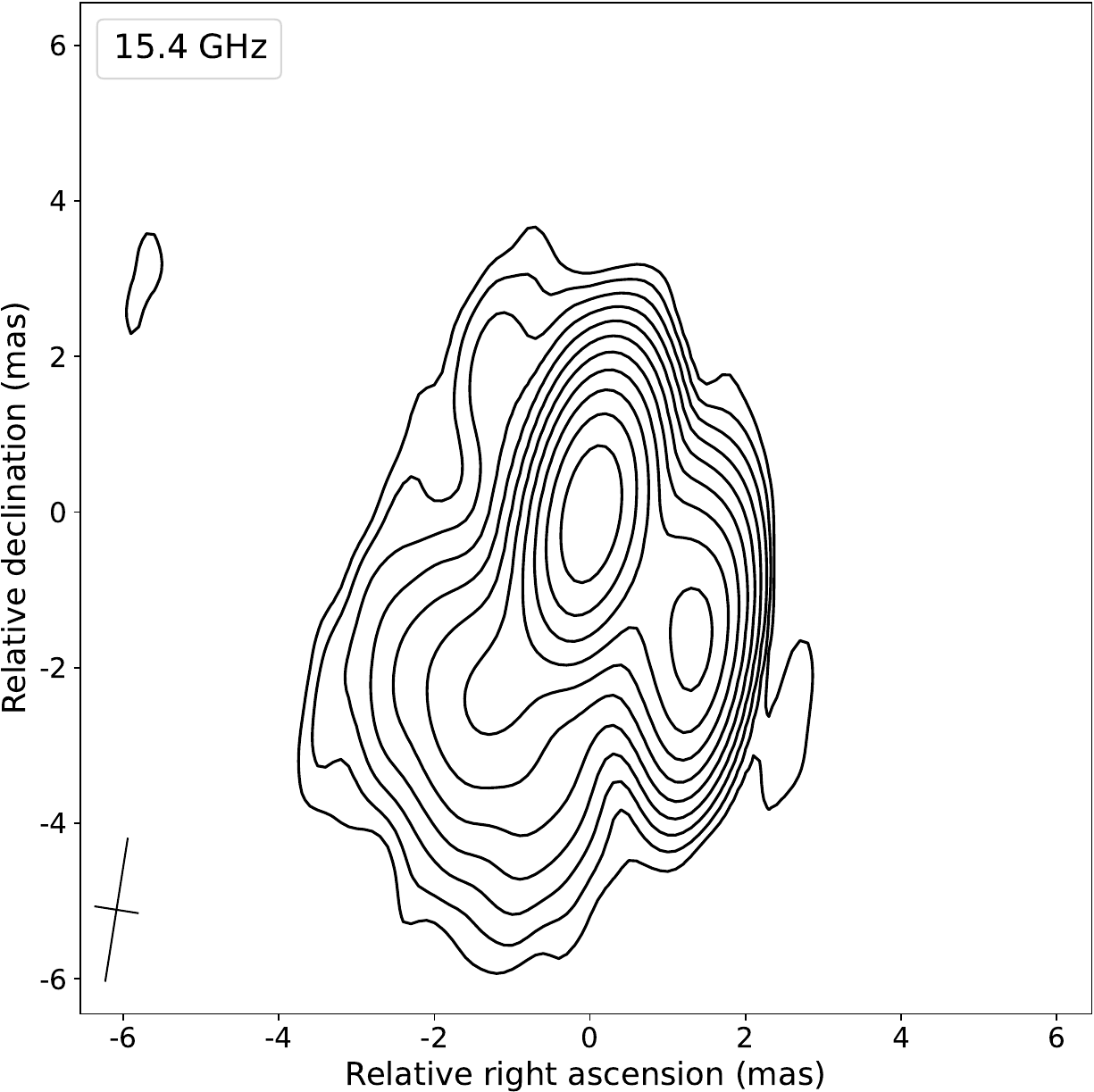}
	\hspace{1cm}
	\includegraphics[width=0.45\linewidth]{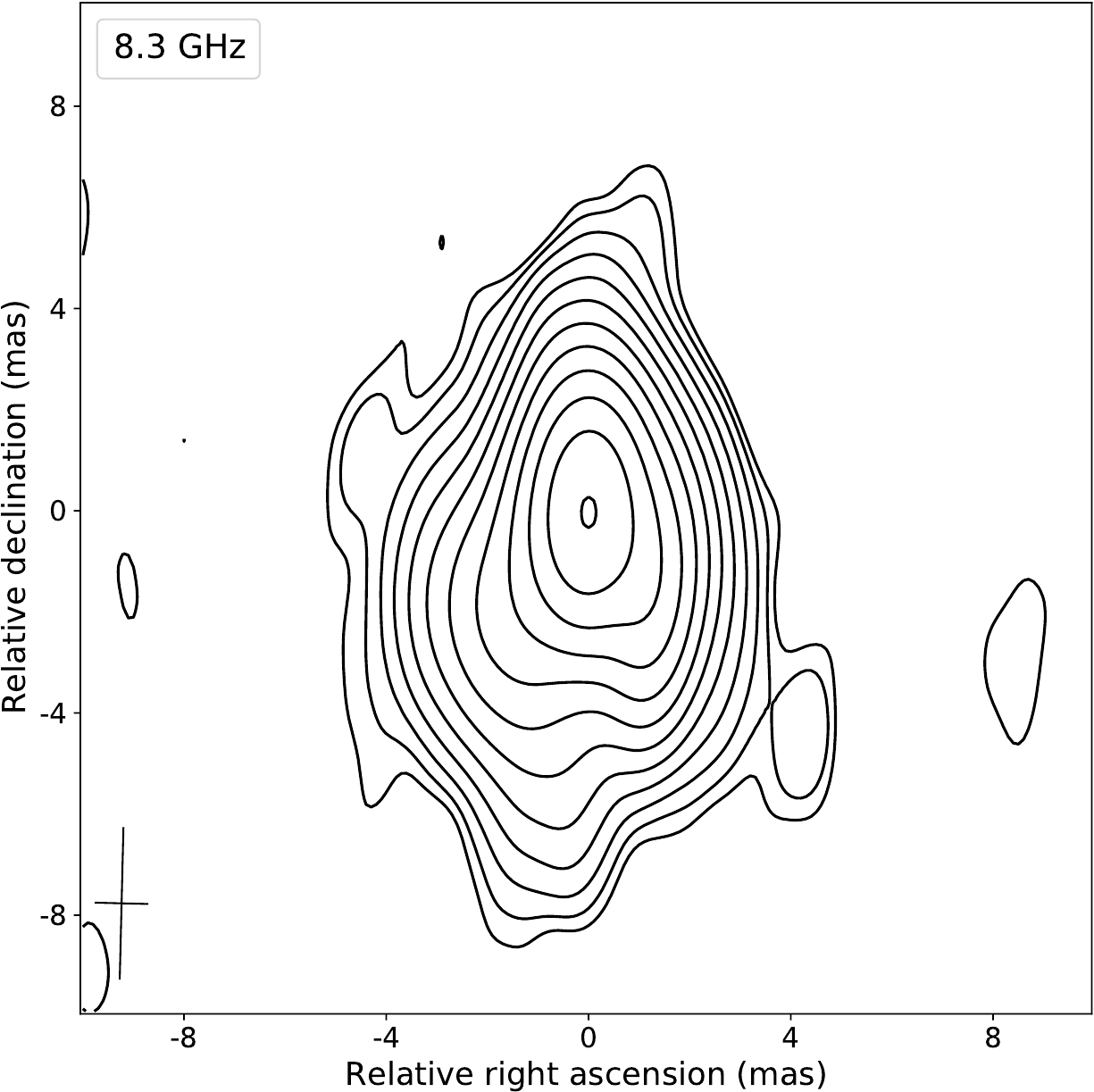}	
    \caption{Stacked total intensity contour maps for quasar 0858$-$279 from the epochs 2005-11-26, 2007-02-09, 2007-05-23, and 2007-08-03 at 15.4 GHz with $\sigma_I = 0.1$ mJy/beam and a map peak of 414 mJy/beam. The averaged beam parameters at this frequency are $B_\mathrm{maj} = 1.8$ mas, $B_\mathrm{min} = 0.6$ mas, and $B_\mathrm{PA} = -8.9$ degrees. For 8.3 GHz, the data from the epochs 2005-11-26, 2007-02-09, 2007-05-23, 2007-08-03, 2017-05-27, 2018-04-08, and 2018-07-21 are used, with $\sigma_I = 0.2$ mJy/beam and a map peak of 583 mJy/beam. The averaged beam parameters at 8.3 GHz are $B_\mathrm{maj} = 3.0$ mas, $B_\mathrm{min} = 1.0$ mas, and $B_\mathrm{PA} = -1.4$ degrees. The intensity contours for both frequencies start at the 2$\sigma_I$ base level and are plotted in $\times2$ steps. The beam, averaged between epochs at each frequency, is shown at the FWHM level in the left bottom corner of the maps. The basic parameters of each map used for stacking are presented in \autoref{tab:intensity_table}}
    \label{fig:stack}
\end{figure*}

\section{VLBA and RATAN-600 data analysis results}
\label{sec:res}

\subsection{Stokes I maps and modelling of the quasar structure}
\label{sec:tot_int_struc}

The general structure of the source in 2007, 2017, and 2018 remained consistent with that of 2005, with no new features appearing in its Stokes I emission. One can find the images at all frequencies from the epoch 2005-11-26 in Fig.~4 in \citetalias[][]{2022MNRAS.510.1480K}.  In \autoref{fig:stack}, we display the stacked Stokes I images of the quasar from all available epochs at 8 and 15 GHz. The resolution of these images was set to an average resolution of all maps used for the reconstruction. We do not show stacked images at 22.2 GHz since there are only three epochs available for stacking, and the resolution is comparable to the resolution of the 8 GHz images. Such low resolution arises because data from some antennas were corrupted. The problems for different time periods included the absence of data, communication problems, bad weather conditions and power glitches (see \autoref{tab:obs} for more detail). Additionally, in \autoref{tab:intensity_table}, we present the main parameters of all maps at all frequencies. 

The stacked image at 15.4 GHz shows an emission in the northeast region relative to the central brightest jet feature. To assess the significance of this radiation, we merged the UV datasets using the DBCON task in AIPS and subsequently reconstructed the source image. This image did not present the signs of the corresponding emission. Thus, we concluded that the radiation does not match the real physical plasma emission from this region.

\begin{table*}
	\centering
	\caption{Main parameters of the Stokes $I$ VLBA maps.}
	\label{tab:intensity_table}
	\begin{tabular}{lrrrcccr}
		\hline\hline
		Epoch & Frequency & Total flux density, $S$ &Map $I$ peak & Map noise, $\sigma_I$ & $B_\mathrm{maj}$ & $B_\mathrm{min}$ & $B_\mathrm{PA}$\\ 
		& (GHz) & (mJy) & (mJy/beam) & (mJy/beam) & (mas) & (mas) & (deg) \\
		\hline
		2007-02-09 & 22.2 & 579 & 411  &  0.8 & 3.2 & 1.1 & $-$4.7 \\   
                   & 15.4 & 727 & 373  &  0.3 & 2.5 & 0.6 & $-$15.4 \\
                   & 8.3 & 1271 & 623  &  0.3 & 3.8 & 1.2 & $-$8.3 \\
                   & 4.8 & 1582 & 725  &  0.8 & 7.2 & 2.2 & $-$6.9   \\  
                   & 2.3 & 1857 & 897  &  1.2 & 23  & 8.6 & 10.6 \\
                   & 1.5 & 1699 & 757  &  2.6 & 46  & 17  & 14.7\\ 
		\hline
		2007-05-23 & 22.2 & 523 & 425  &  1.0 & 4.2 & 1.6 & 21.1 \\   
                   & 15.4 & 792 & 389  &  0.2 & 1.6 & 0.5 & $-$8.3 \\
                   & 8.3 & 1275 & 662  &  0.3 & 2.7 & 1.0 & $-$2.4 \\
                   & 4.8 & 1758 & 728  &  0.4 & 4.8 & 1.7 & $-$0.3   \\  
                   & 2.3 & 2256 & 981  &  0.8 & 19.4& 6.3 & 19.8 \\
                   & 1.5 & 1752 & 714  &  1.8 & 37  & 13.5  & 14.1\\ 
		\hline
		2007-08-03 & 15.4 & 784 & 391  &  0.2 & 1.5 & 0.6 & $-$4.5 \\
                   & 8.3 & 1299 & 713  &  0.2 & 3.3 & 1.2 & $-$5.9 \\
                   & 4.8 & 1795 & 921  &  0.6 & 5.8 & 2.3 & $-$2.1 \\  
                   & 2.3 & 2018 & 1100 &  1.4 & 25  & 9.5 & 6.5 \\
                   & 1.5 & 1712 & 744  &  1.2 & 43  & 16  & 11.8\\ 
		\hline
		2017-05-27 & 8.3  & 981 & 428  &  0.5 & 2.9 & 0.9 &  5.9 \\
		\hline
		2018-04-08 & 8.3  & 1049 & 393 &  0.5 & 2.0 & 0.8 & $-$0.1 \\
		\hline
		2017-07-21 & 8.3 & 1037 & 450  &  0.3 & 2.4 & 0.9 & 0.7 \\
	    \hline
	\end{tabular}
	\begin{tablenotes}
            \item \emph{Note.} Total flux density $S$ was calculated as a sum of CLEAN components. $B_\mathrm{maj}$ and $B_\mathrm{min}$ are the major and minor axis full width at half maximum (FWHM) of the restoring beam, respectively. $B_\mathrm{PA}$ is the position angle of the beam. We do not show the parameters of the 22.2 GHz map from the epoch 2007-08-03 since we were not able to reconstruct the image at this frequency due to problems with the data (see \autoref{sec:data} and \autoref{tab:obs}).
        \end{tablenotes}
\end{table*}

In our earlier analysis, we determined that the southwest region was associated with the opaque core, and the prominent central feature was actually a part of the jet. Lower frequencies are not discussed in this analysis due to their limited resolution, which made detailed structural study challenging. To study the structure and its evolution in more detail, we fitted circular Gaussian components to the visibility data in Difmap with the modelfit independently for different epochs. We continued to add new Gaussian components only when they led to a significant reduction in $\chi^2$. Thus, this approach enabled us to find the optimal minimum number of components.  In \autoref{tab:gauss_comp}, we give their main parameters at 8.3, 15.4 and 22.2 GHz for all epochs. Typical errors turned out to be 5-10$\,\%$. In addition, \autoref{fig:components} shows the results of our modelling at 15.4 GHz. We note that we were able to model the structure with a minimum of four components at 15.4 GHz. Meanwhile, for the 2007, 2017, and 2018 epochs, three components best represented the data at 8.3 GHz. We tried to distinguish between the core component C and the next one along the jet, S, further downstream, by fitting additional delta components (also referred to as CLEAN components) in this region, alongside the Gaussian components. However, the fit did not converge due to low resolution.

A notable observation is that the number of Gaussian components best describing the brightest jet feature transitioned from two (J$_0$ and J$_1$ components) to just one (J component) over a year-long period. On the one hand, this might be directly related to the evolution of a shock wave across that specific part of the jet (see \autoref{sec:discus} for more detail). On the other hand, it can be an artificial effect since the fitted components do not always represent real physical features in the jet \citep[e.g.][]{2009AJ....138.1874L}.

\begin{table}
	\centering
    \caption{Circular Gaussian model fitting results. 
    \label{tab:gauss_comp}}	
    \begin{threeparttable}
    \resizebox{1\linewidth}{!}{
	    \begin{tabular}{ccccrcc} 
		    \hline\hline
            Epoch & Frequency & Flux  & Distance  &  P.A. &  FWHM & Component \\
             &  & density  &  & & & label \\
             & (GHz) & (Jy) & (mas) & (deg.) & (mas) &\\
            \hline
            \hline
            2005-11-26 & 22.2  & 0.07  &  2.46  &  $-$143 & 0.22 & C  \ \ \\   
                       & & 0.44  &  0.05  &  139    & 0.35& J$_0$\\   
                       & & 0.07  &  0.34  &  $-$91  & 0.21 & J$_1$\\
                       & & 0.03  &  2.49  &  156    & 1.09 &\\  
                       & & 0.04  &  1.35  &  $-$137 & 0.72 & S \ \ \\
             
             \hline
                        & 15.4 & 0.06  &  2.35  &  $-$143 & 0.37& C \ \ \\ 
                        &  & 0.59  &  0.06  &  90     & 0.44& J$_0$\\   
                        &  & 0.15  &  0.29  &  $-$78  & 0.39& J$_1$\\
                        &  & 0.08  &  2.36  &  156    & 1.72 &\\
                        &  & 0.07  &  1.60  &  $-$142 & 0.66 & S \ \ \\
            \hline
                        &8.3 &0.19  &  1.80  &  $-$143 & 1.25 & C \ \ \\ 
                        &    &1.12  &  0.05  &  $-$17  & 1.04 &J \ \ \\   
                        &    &0.09  &  1.04  &  155    & 0.84 &\\
                        &    &0.14  &  2.68  &  148    & 1.80 &\\
            \hline
            \hline
             2007-02-09 & 22.2 & 0.09  &  2.14  &  $-$143 & 0.53& C \ \ \\ 
                        &  & 0.45 &  0.07  &  21  & 0.54 & J \ \ \\   
                        &  & 0.03  &  1.67  &  121    & 1.03 &\\
            \hline
            
                        & 15.4 & 0.08  &  2.21  &  $-$143 & 1.25& C \ \ \\ 
                        &  & 0.53 &  0.03  &  134  & 0.61 & J \ \ \\   
                        &  & 0.07  &  2.50  &  151    & 1.57 &\\
                        &  & 0.05  &  1.21  &  $-$136    & 0.79 & S \ \ \\
            \hline
                        &8.3 &0.13  &  1.67  &  $-$124 & 0.83& C \ \ \\ 
                        &    &0.97  &  0.03  &  $-$19  & 1.48 & J \ \ \\   
                        &    &0.16  &  2.28  &  141    & 2.07 &\\
                       
            \hline
            \hline
            2007-05-23 & 22.2 & 0.06  &  2.10  &  $-$140 & 0.11& C \ \ \\ 
                        &  & 0.39 &  0.06  &  59  & 0.37 & J \ \ \\   
                        &  & 0.07  &  1.63  &  154    & 2.49 &\\
            \hline
                       & 15.4 & 0.08  &  2.23  &  $-$143 & 0.30& C \ \ \\ 
                       &  & 0.59 &  0.02  &  30  & 0.53 & J \ \ \\   
                       &  & 0.07  &  2.69  &  152    & 1.91 \\
                       &  & 0.05  &  1.55  &  $-$141   & 0.48 & S \ \ \\
            \hline
                        &8.3 &0.14  &  1.88  &  $-$141 & 0.79& C \ \ \\ 
                        &    &1.01  &  0.04  &  $-$29  & 1.08 & J \ \ \\   
                        &    &0.13  &  2.53  &  142    & 1.63 &\\

            \hline
            \hline
            2007-08-03 & 15.4 & 0.09  &  2.21  &  $-$143 & 0.33& C \ \ \\ 
                       &  & 0.55 &  0.01  &  50  & 0.50 & J \ \ \\   
                       &  & 0.09  &  2.31  &  155    & 2.03 &\\
                       &  & 0.06 &  1.53  &  $-$140    & 0.50 & S \ \ \\
            \hline
                        &8.3 &0.18  &  1.90  &  $-$138 & 1.13& C \ \ \\ 
                        &    &0.98  &  0.09  &  20  & 1.06 & J \ \ \\   
                        &    &0.13  &  2.54  &  146 & 1.85 &\\
            \hline
            \hline
            2017-05-27 & 8.3 & 0.26  &  1.77  &  $-$147 & 0.82& C \ \ \\ 
                       &  & 0.59 &  0.09  &  52  & 0.91 & J \ \ \\   
                       &  & 0.10  &  2.90 &  143    & 1.72 &\\
            \hline
            \hline   
            2018-04-08 & 8.3 & 0.30  &  2.08  &  $-$149 & 0.72&C \ \ \\ 
                       &  & 0.60 &  0.06  &  51  & 0.79 & J \ \ \\   
                       &  & 0.12  &  1.89  &  126   & 1.38 &\\
            \hline
            \hline   
            2018-07-21 & 8.3 & 0.16  &  2.11  &  $-$144 & 0.62& C \ \ \\ 
                       &  & 0.70 &  0.08  &  33  & 0.88 & J \ \ \\   
                       &  & 0.14  &  2.17  &  134    & 1.32 &\\
            \hline
            \hline   
            
        \end{tabular}
        }

        \begin{tablenotes}
            \begin{minipage}{8cm}
            \item \emph{Note.} The table shows: (1) the corresponding epoch, (2)  the frequency of observation of the quasar, (3) the flux density from the component, (4) the radial distance of the component centre from the centre of the map, (5) the position angle of the centre of the component, (6) the FWHM size of the circular Gaussian component, (7) the component label. 'C' denotes the component associated with the core, 'J', 'J$_0$', and 'J$_1$' represent the dominant jet components, and 'S' is the component situated between the core and the jet components. Refer to \autoref{fig:components} for more details.
            \end{minipage}

        \end{tablenotes}
	\end{threeparttable}
\end{table}

\subsection{Kinematics}
\label{sec:kinem}

\begin{figure}
	\includegraphics[width=\linewidth,trim={1cm 0cm 1cm 1cm},clip]{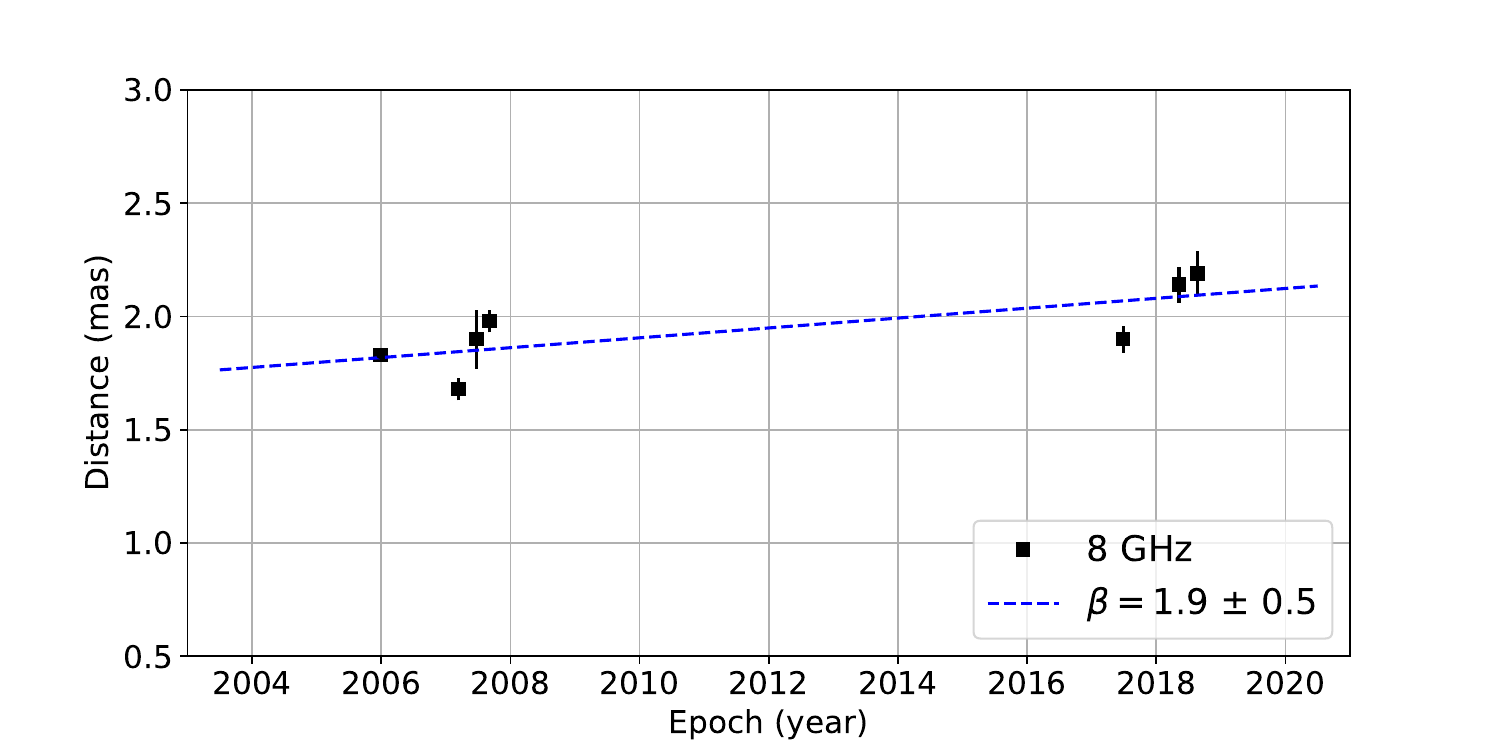}
    \caption{Evolution of the distance between the core component C and the jet feature component J using the 8.3 GHz data. Time is represented in years. The blue dashed line corresponds to a linear fit used to determine the apparent speed of the component J.}
    \label{fig:kinem}
\end{figure}

We analyzed the variation in distance between the core component C and the jet component J over time, as illustrated in \autoref{fig:kinem}, utilizing the 8.3 GHz data which encompasses all seven epochs. The results show that the distance increased with the jet feature apparent speed of  $(1.9 \pm 0.5)c$ assuming that the core component is stationary. Such behaviour might hint at how the component slowly carves its way through the external medium. If we assume that the jet is observed at critical angle \citep[e.g.][]{2021ApJ...923...67H}, the relativistic Doppler factor can be estimated as $\delta = \sqrt{1+\beta^2} \sim \beta \sim 2$. This is consistent with the results obtained in \citetalias[][]{2022MNRAS.510.1480K}, where the variability Doppler factor was estimated. This result, however, does not conclusively rule out the possibility of a stationary feature where the apparent motion is caused by changing opacity effects in the source.

\subsection{Variability of the parsec-scale jet structure}
\label{sec:vlbi_var}

To study the variability properties of the quasar, we need to align the images from different epochs. That was not a trivial process since the absolute positions of sources in the sky are lost during the self-calibration process. Apart from that, different components can evolve between multi-epoch observations, and new ones can emerge due to the ejection of plasma material. Finally, variable synchrotron self-absorption can have an impact on visible positions. As a consequence, this ambiguity in positions and the number of components complicates the process of component identification across epochs. However, modelling of the quasar structure demonstrated that the core component C was consistently visible across all epochs (see also spectral index maps in \autoref{sec:spectral}).

\begin{figure*}
	\includegraphics[width=0.45\linewidth,trim={0cm 0cm 0cm 0cm},clip]{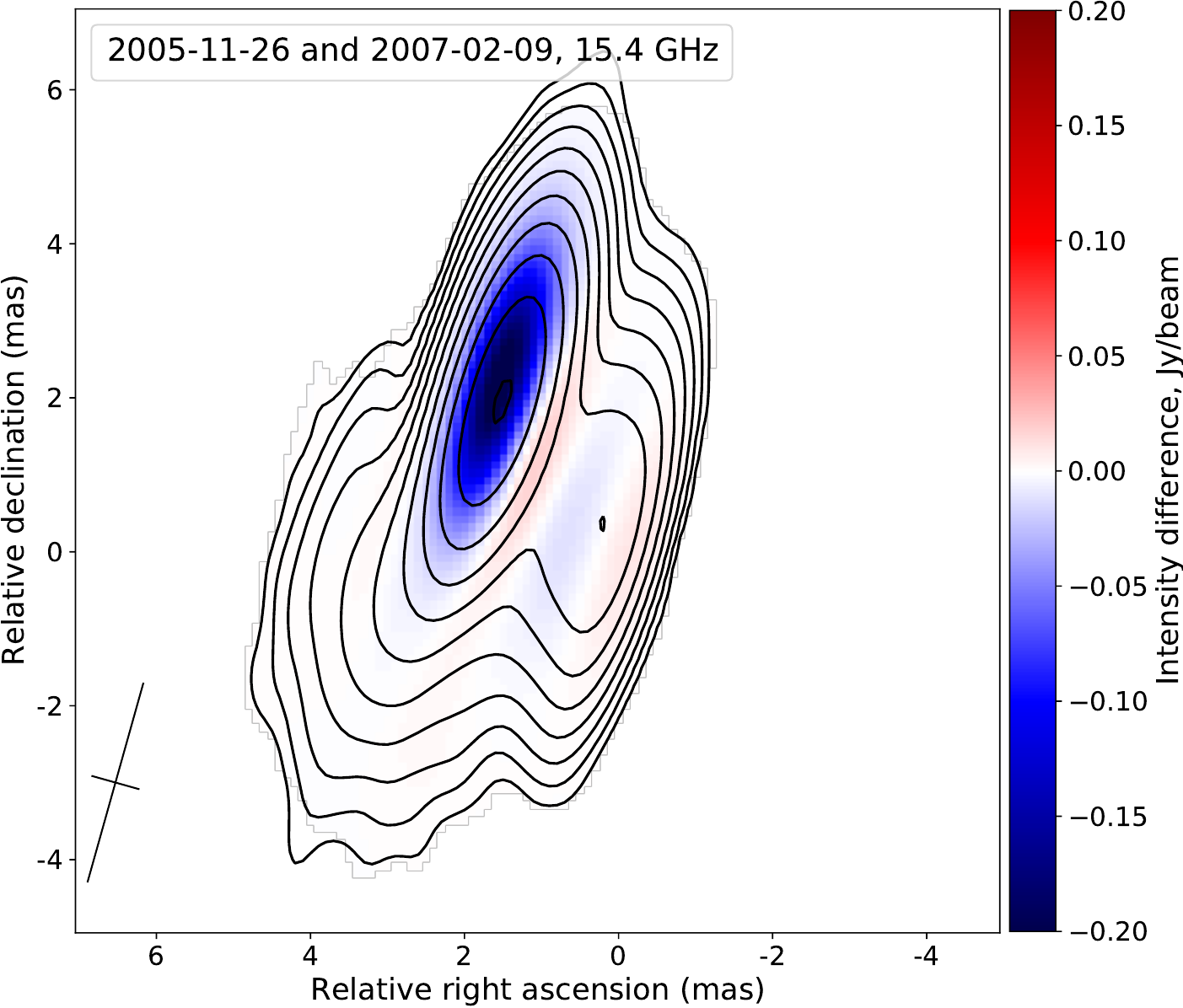}
	\hspace{1cm}
	\vspace{0.5cm}
	\includegraphics[width=0.45\linewidth,trim={0cm 0cm 0cm 0cm},clip]{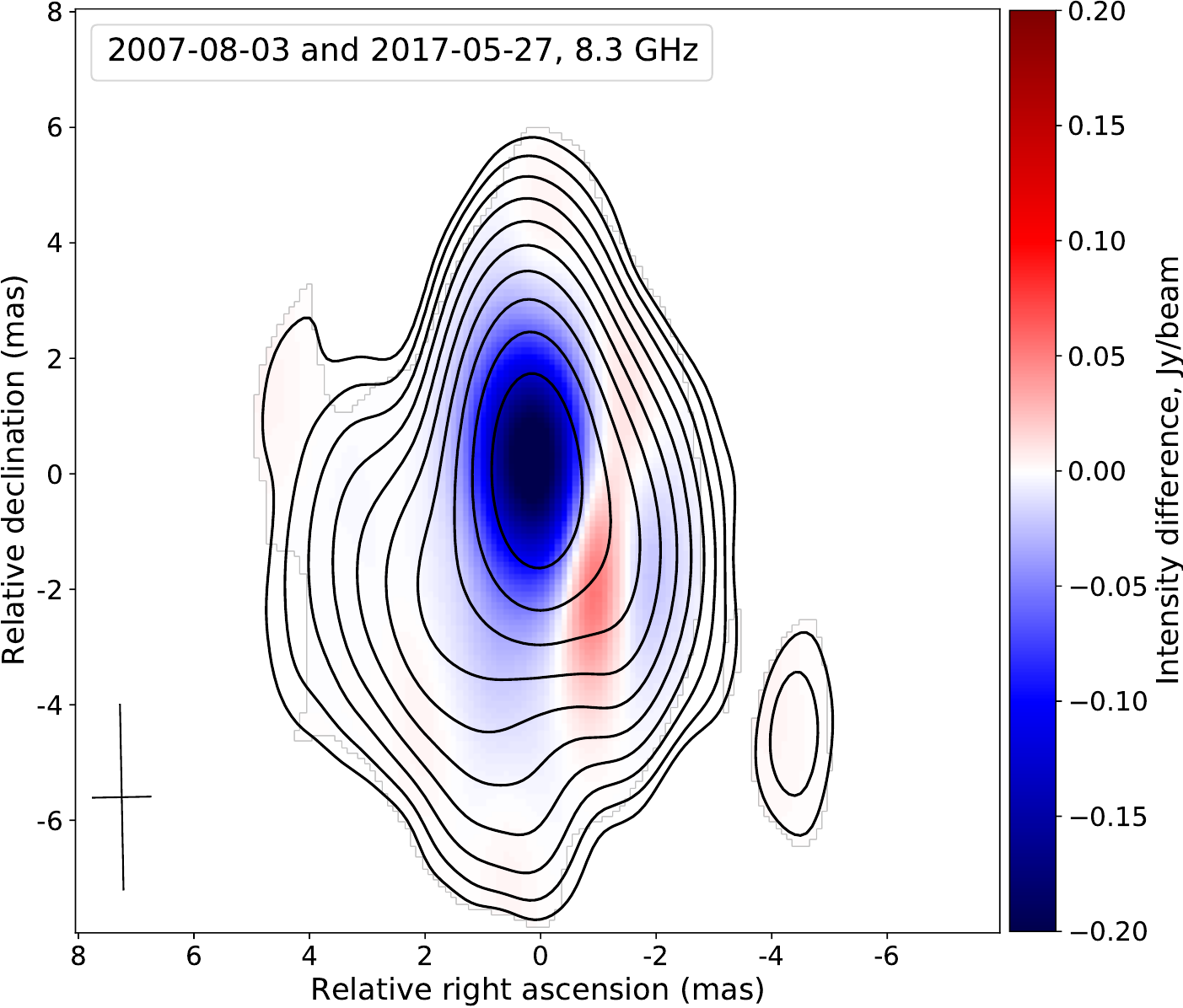}
	
	\caption{Difference between the Stokes I maps, specifically showing the later epoch minus the earlier epoch as indicated in the legend at the corresponding frequency. The total intensity contour maps are taken from the earliest corresponding epoch. The contours start at the 2$\sigma_I$ base level and are plotted in $\times2$~steps. The averaged between epochs beam is shown at the FWHM level in the left bottom corner. The intensity difference values are shown only in those pixels where intensity exceeds 2$\sigma_I$. We show images for these specific epochs since they demonstrate significant changes in Stokes I emission. The image on the left has $\sigma_I = 0.3$ mJy/beam and a map peak of 569 mJy/beam. The averaged beam parameters are $B_\mathrm{maj} = 2.7$ mas, $B_\mathrm{min} = 0.6$ mas, and $B_\mathrm{PA} = -15.7$ degrees. The image on the right has $\sigma_I = 0.3$ mJy/beam and a map peak of 663 mJy/beam. The averaged beam parameters are $B_\mathrm{maj} = 3.2$ mas, $B_\mathrm{min} = 1.0$ mas, and $B_\mathrm{PA} = 1.1$ degrees.}
    \label{fig:int_dif}
\end{figure*}

We tried several approaches to align the images at 15.4 GHz firstly between 2005-11-26 and 2007-02-09. They included aligning based on the position of the core component as well as by the positions of the J and J$_0$ or J and J$_1$ components. After visually inspecting the differences between the Stokes I images, we determined that aligning based on the core component yielded the best results. In \autoref{fig:int_dif}, we present the difference between the Stokes I images of the quasar at 15.4 GHz, as reconstructed with the same pixel sizes, uv-coverage and an epoch averaged beam. At the same time, other alignment methods produced results with minor artefacts in the image, but these did not significantly deviate from the results obtained by aligning with the core component.

The most drastic changes in the 2005 and 2007 epochs occurred in the brightest jet feature. The changes in the core's flux density and in the southeast component were almost negligible. Additionally, the changes in the emission between three epochs in 2007 turned out to be minor, showing that the flux density was basically unaltered both in the core region and in the jet. Therefore, we do not show these images.

We investigated the variation in radiation between 2007, 2017, and 2018 by aligning the Stokes I images at 8.3 GHz. The best results for the intensity difference maps were obtained by the jet feature alignment. We present them in \autoref{fig:int_dif}. The core alignment yielded similar findings, meaning that in both cases, the most drastic changes were observed mainly in the jet feature. However, the core alignment resulted in a difference image where the region of flux change was not as symmetric relative to the center of the component as it was in the jet alignment. We note that the flux density dropped in the jet feature J between 2007 and 2017. There was a small flare in 2018 with flux density rising in both core and jet compared to 2017. In accordance with the causality principle, the simultaneous growth in these two regions cannot be linked. The projected distance between the core and the component J is approximately 2 mas, which is about 17 pc. Such a distance makes it improbable that the flare had time to reach the component J. Finally, the core flux started to drop in the same year, while the jet flux density increased slightly. Due to the minimal impact of these changes, we have chosen not to display them. \autoref{tab:gauss_comp} shows more detail on the amount of flux density changes in the core and the jet features.   

\subsection{RATAN-600 single-dish spectrum variability}
\label{sec:ratan_var}

\begin{figure*}
	\includegraphics[width=0.94\linewidth]{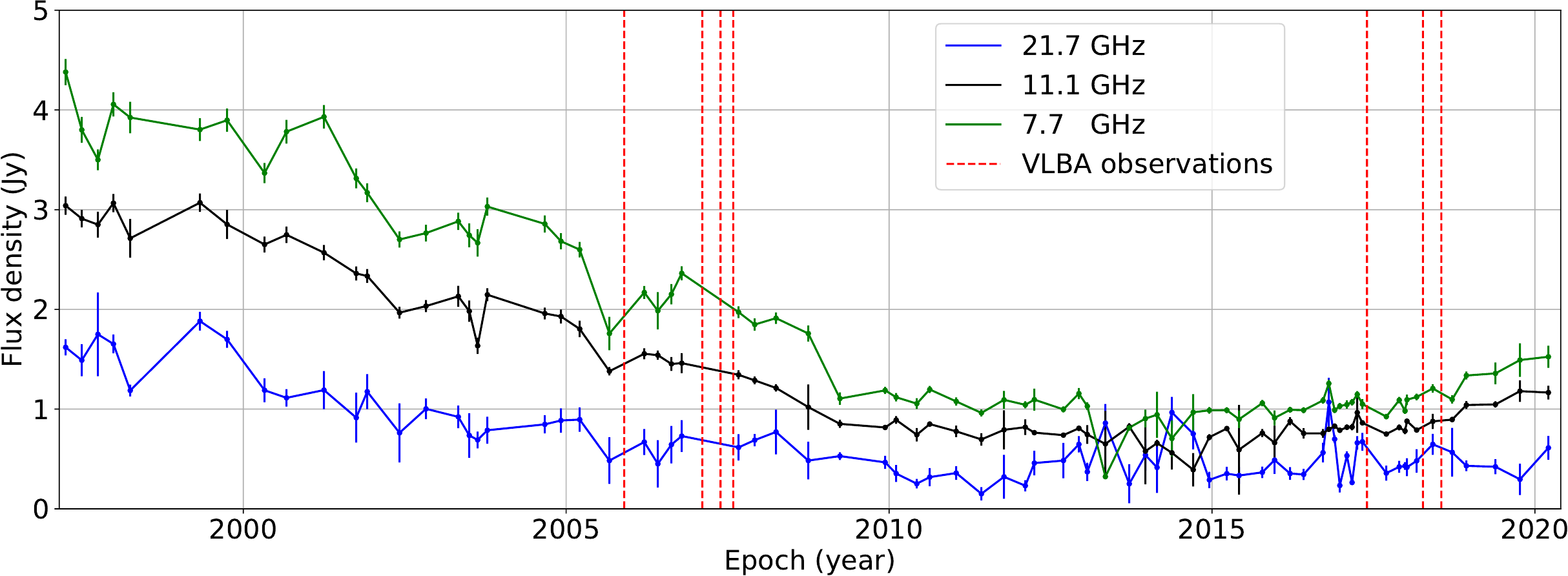}
    \caption{Multi-frequency RATAN-600 light curve of the GPS quasar 0858$-$279 for three frequencies. The red dashed lines show the epochs of the VLBA observations.}
    \label{fig:light_curve}
\end{figure*}

We conducted a long-term variability analysis of the quasar using RATAN-600 data. Light curves for the three highest frequencies are shown in \autoref{fig:light_curve}. We modeled the integrated spectrum of the quasar at each epoch using the formula that describes the spectrum of a synchrotron homogeneous source \citep[e.g.][]{1966Natur.211.1131V,1970ranp.book.....P}:
\begin{equation}
    I(\nu) = I_{\mathrm{m}}\left ( \frac{\nu}{\nu_{\mathrm{m}}} \right )^{2.5}\frac{1-\mathrm{exp}(-\tau_{\mathrm{m}}(\nu/\nu_{\mathrm{m}})^{\alpha_{\mathrm{thin}}-2.5})}{1-\mathrm{exp}(-\tau_{\mathrm{m}})}\,,
	\label{eq:homog_synchr_rad_eq1}
\end{equation}
where $I_{\mathrm{m}}$ is the maximum intensity in the spectrum reached at the corresponding frequency $\nu_{\mathrm{m}}$, and $\tau_{\mathrm{m}}$, the optical depth at the turnover frequency $\nu_{\mathrm{m}}$, is approximated as:
\begin{equation}
    \tau_{\mathrm{m}} = \frac{3}{2} \left( \sqrt{1 - \frac{8\alpha_{\mathrm{thin}}}{3 \times 2.5}} - 1 \right).
\end{equation}

\begin{figure}
	\includegraphics[width=1.\linewidth,trim={0cm 0cm 0cm 0cm},clip]{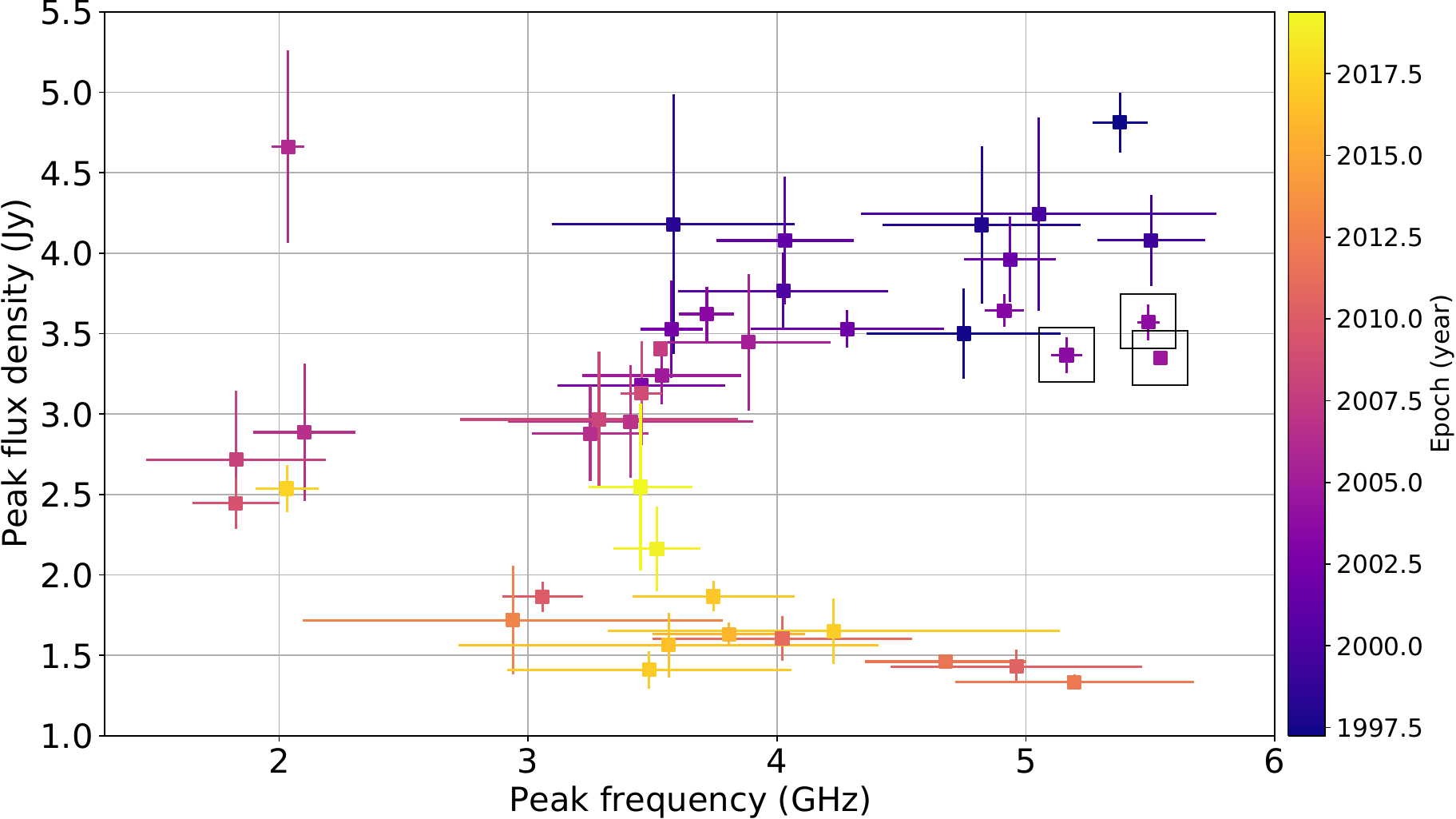}
	\caption{Evolution of the turnover frequency and the peak flux density obtained from the RATAN-600 data on the ($\nu_\mathrm{m}$ $-$ $S_\mathrm{m}$)-plane. The colours show the corresponding epoch. The points in the black squares displaced to the right do not follow the C-shaped evolution track. Such behaviour could be related to the shock wave interaction (see \autoref{sec:discus} for more detail). }
    \label{fig:peak_fl_fr}
\end{figure}

In \autoref{fig:peak_fl_fr}, we illustrate the evolution of these parameters on the ($\nu_\mathrm{m}$,$S_\mathrm{m}$)-plane. The errors were derived using the variance–covariance matrix of the fit. We notice that at some epochs the fit did not converge. Therefore, the number of points in \autoref{fig:peak_fl_fr} is lower than the total number of epochs in the RATAN-600 observations. We acknowledge that the assumption of homogeneity for the integrated spectrum is somewhat simplistic, given that the spectral distribution includes contributions from various jet regions. However, we utilised this conservative approach in further analysis instead of the generalised one with a varying index of optically thick radiation (see eq. (7) in \citetalias[][]{2022MNRAS.510.1480K}) to keep the number of free parameters to a minimum. Apart from that, the generalised approach showed the same evolutionary trends for the evolution of $I_{\mathrm{m}}$ and $\nu_{\mathrm{m}}$. 

\subsection{Spectral index maps}
\label{sec:spectral}

\begin{figure*}
	\includegraphics[width=0.46\linewidth,trim={0cm 0cm 0cm 0cm},clip]{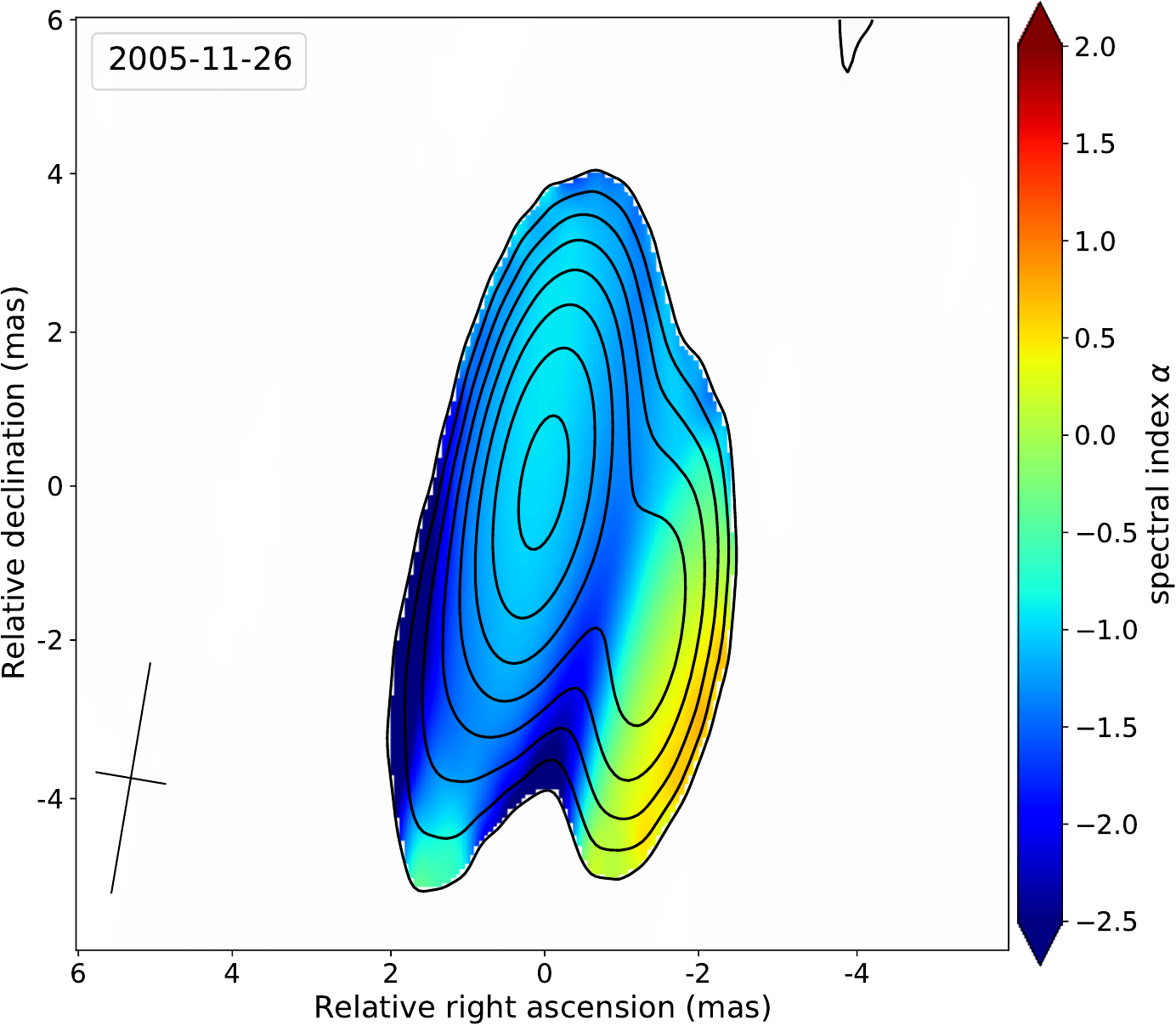}
	\hspace{0.6cm}
	\includegraphics[width=0.46\linewidth,trim={0cm 0cm 0cm 0cm},clip]{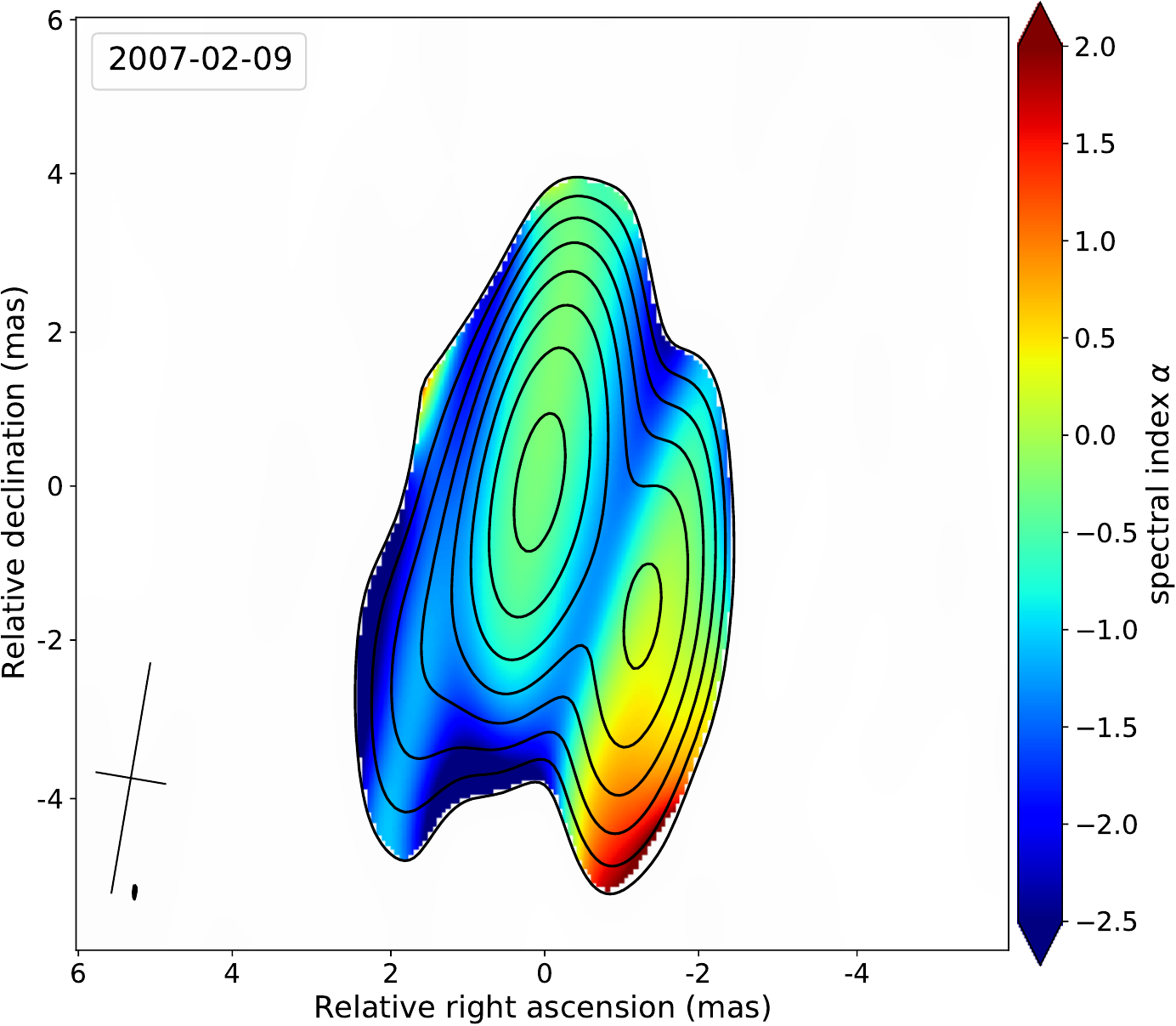}
	
	\includegraphics[width=0.46\linewidth,trim={0cm 0cm 0cm 0cm},clip]{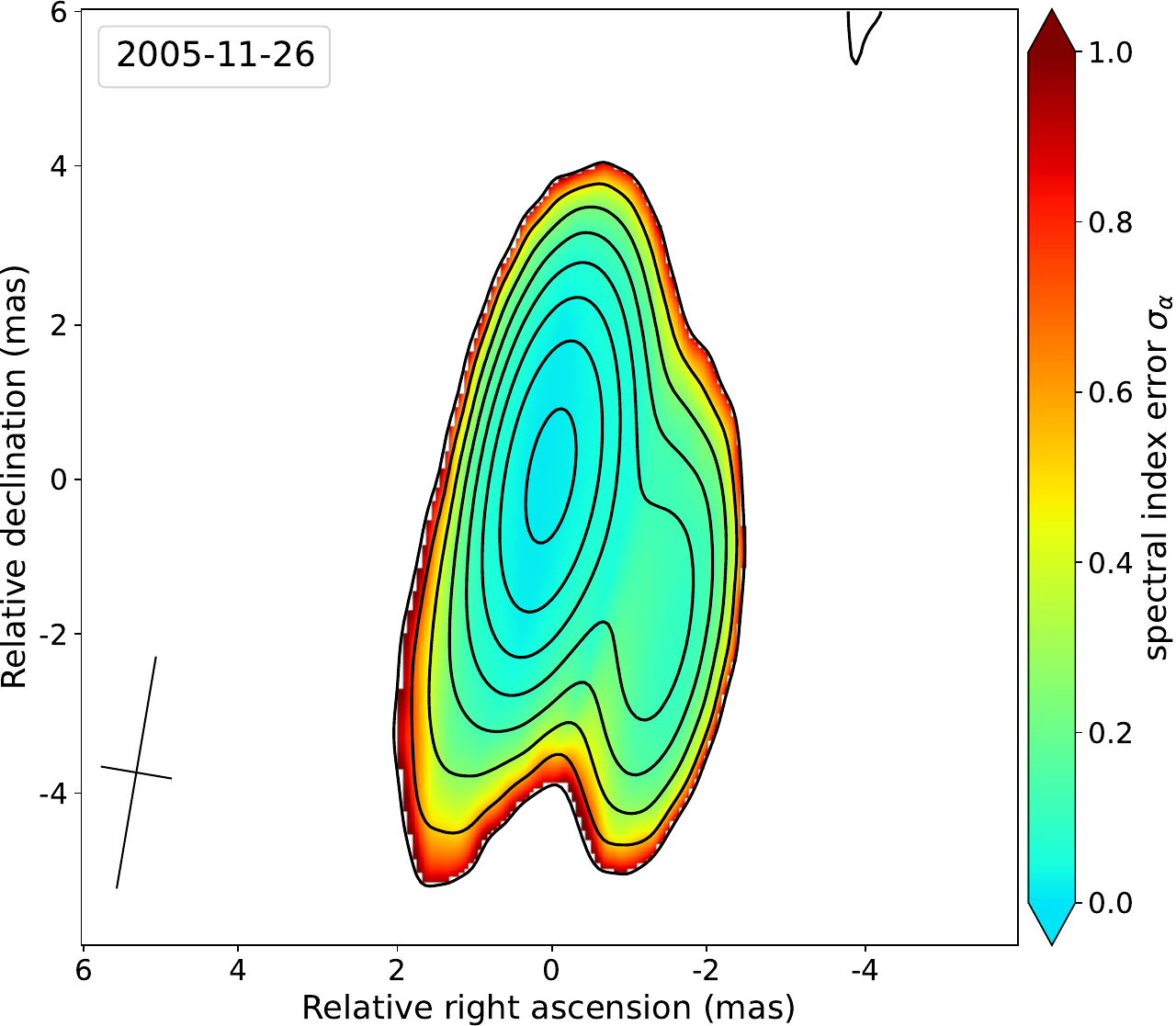}
	\hspace{0.6cm}
	\includegraphics[width=0.46\linewidth,trim={0cm 0cm 0cm 0cm},clip]{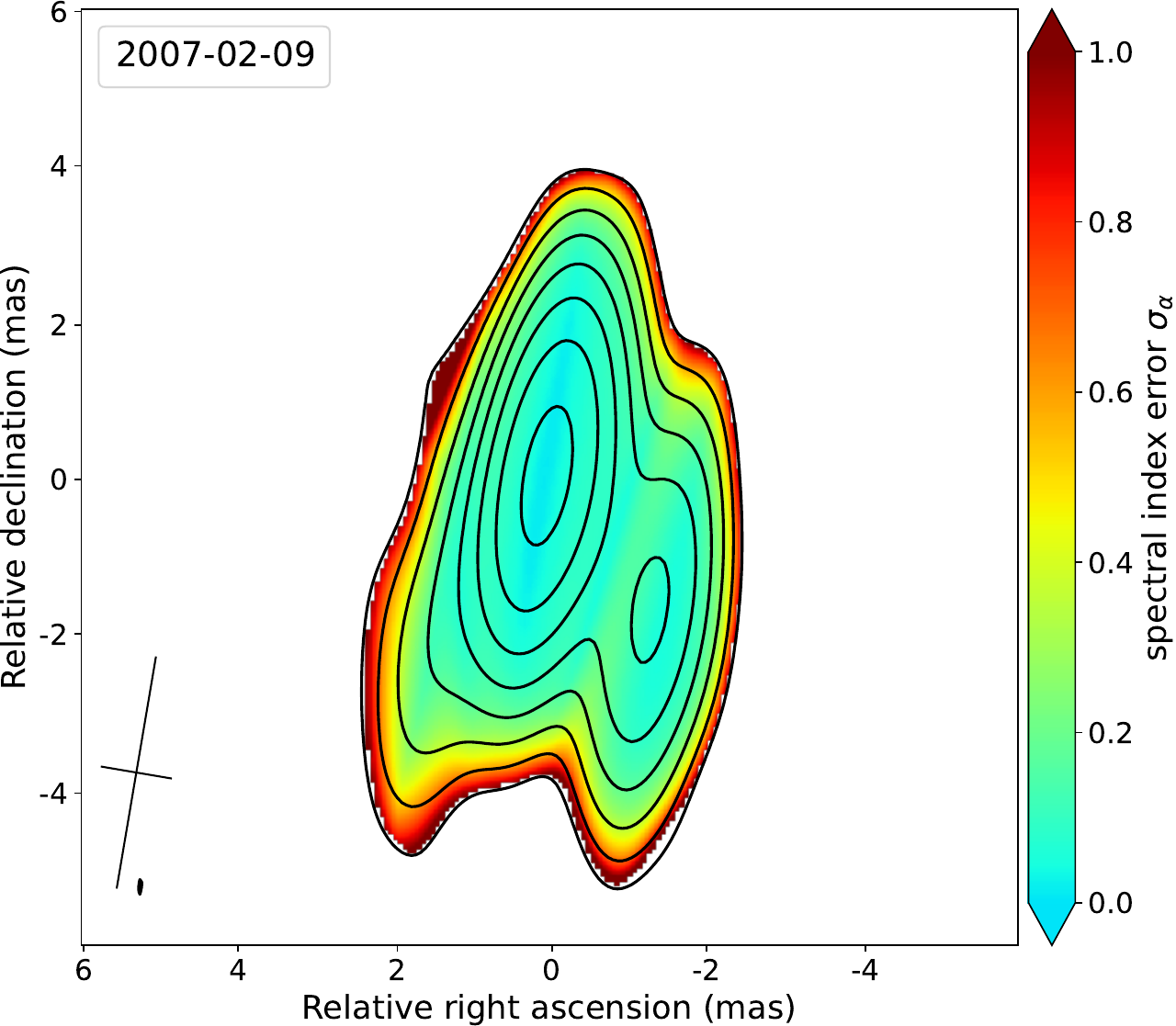}
	
	\caption{Spectral index (top) and spectral index error (bottom) maps from the epoch 2005-11-26 (left) and 2007-02-09 (right) for 15$-$22~GHz frequency intervals. The spectral index is defined as $S \propto \nu^{+\alpha}$. The spectral index error maps represent the one sigma error of the spectral index (see also eq. (4) in \citetalias{2022MNRAS.510.1480K}). The maps show full intensity contours starting at the 3$\sigma_I$ level taken from the Stokes $I$ map for 22.2~GHz. The spectral index and the spectral index error are presented only in the pixels where intensity exceeds 3$\sigma_I$. The maps are reconstructed with the same average beam taken from the epoch 2007-02-09, which is shown in the bottom left corner at the FWHM level. The beam parameters are $B_\mathrm{maj} = 3.0$~mas, $B_\mathrm{min} = 0.9$~mas, and $B_\mathrm{PA} = -9.6$~deg. The images on the left have a $\sigma_I$ of 0.9 mJy/beam and a map peak of 453~mJy/beam. The images on the right have a $\sigma_I$ of 0.8~mJy/beam and a map peak of 400~mJy/beam. The errors are also discussed in \autoref{sec:spectral}.}
    \label{fig:sp_ind}
\end{figure*}

Throughout this paper, we define the spectral index $\alpha$ as $S \propto \nu^{+\alpha}$, where $S$ is the flux density observed at frequency $\nu$. This convention is used for all spectral index calculations and discussions.
\autoref{fig:sp_ind} displays the reconstructed spectral index and the corresponding error maps for the 15-22 GHz range during two epochs: 2005-11-26 and 2007-02-09, both with identical angular resolution. We aligned the images at different frequencies by using the optically thin jet components \citep[e.g.][]{2008A&A...483..759K,2011A&A...532A..38S,2013A&A...557A.105F} as a reference similar to the 2005 epoch analysis, because their absolute positions do not depend on frequency. The derived spectral index maps exhibit a similar structure with a clearly visible opaque core in the southwest region. Additionally, the values increase in the core region compared with the 2005-11-26 epoch. This could be partially caused by boundary effects, which refer to the increased uncertainty in areas where the signal-to-noise ratio is lower. To test if this increase is significant, we calculate spectral indices for the corresponding core components taken from \autoref{tab:gauss_comp}. The core spectral index value in 2005 turns out to be $0.4 \pm 0.4$, while in 2007, it is $1.1 \pm 0.4$. We conclude, therefore, that the increase is significant.  
Apart from that, the spectral indices also increase in the brightest jet feature. This can be related to the opacity properties, which are critical in the subsequent analysis. This is considered carefully when studying the polarization properties of the source in \autoref{sec:pol_prop}. The spectral index error is calculated in the same manner as in \citetalias[][]{2022MNRAS.510.1480K}. Owing to the low resolution at 22 GHz and the varying opacity effects at 8 GHz for the 2007 epochs (\autoref{sec:vlba_obs}), we did not measure the core shift \citep{1979ApJ...232...34B,1981ApJ...243..700K,1998A&A...330...79L,2012A&A...545A.113P,2019MNRAS.485.1822P}, and as a result, magnetic field estimates for the core region during these epochs were not determined.

\subsection{Polarization properties}
\label{sec:pol_prop}

\begin{figure*}
	\includegraphics[width=1\linewidth,trim={0cm 0cm 0cm 0cm},clip]{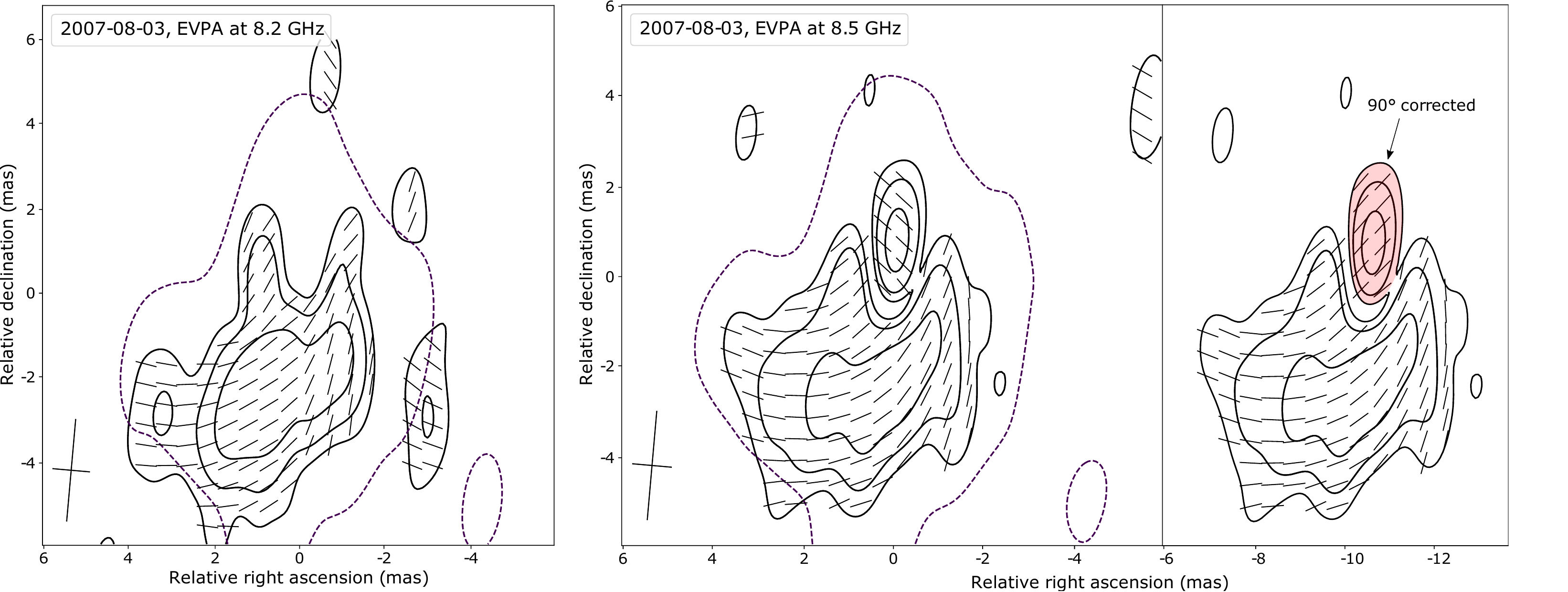}
	\caption{EVPA map not corrected for Faraday rotation at 8.2~GHz and 8.5~GHz. The maps show the outer contour (dotted line) taken from the maps of the total intensity at the 5$\sigma_I$ level.
    The EVPA direction is shown as black strokes within the linear polarization contours starting at the 3$\sigma_P$ level with the $\times2$ steps. 
    The same linear polarization contours with corrected for the opacity effects EVPAs for 8.5 GHz (see details in \autoref{sec:pol_prop}) are shown on the right. The part of the structure with EVPAs which were corrected by 90$^\circ$ to obtain RM maps is highlighted with the transparent red colour. The restoring beam is shown as a cross in the bottom left corner at the FWHM level. For the 8.2 GHz map, $\sigma_I$ is 0.3 mJy/beam with a peak intensity of 591 mJy/beam, and a peak in linear polarization ($P$ peak) of 6.1 mJy with $\sigma_P$ of 0.3 mJy/beam. The beam parameters are $B_\mathrm{maj} = 2.4$ mas, $B_\mathrm{min} = 0.9$ mas, and $B_\mathrm{PA} = -5.0$ degrees. For the 8.5 GHz map, $\sigma_I$ is 0.3 mJy/beam with a peak intensity of 587 mJy/beam, and a $P$ peak of 5.1 mJy with $\sigma_P$ also of 0.3 mJy/beam. The beam parameters for this map are the same as for 8.2 GHz.}
    \label{fig:evpa_map}
\end{figure*}

We performed polarization analysis for the three epochs in 2007. First, we reconstructed fractional polarization images. They do not demonstrate any significant changes in the structure compared to the 2005 epoch, but, over time,  the radiation becomes systematically less polarized at the highest frequencies. \autoref{tab:lin_degree_table} shows the degree of linear polarization for the dominant component J. Second, some of the EVPA images showed clear presence of opacity effects. For example, in \autoref{fig:evpa_map}, a distinct region in the center of the 8.5~GHz map has EVPAs perpendicular to those in the surrounding areas. This can be explained in terms of an emission regime which becomes optically thick at around 8~GHz in the centre region, while being optically thin in the outer parts. The evidence for opacity in this central part is a low degree of polarization and high spectral index. This effect is stable across all epochs in 2007 and the 2005 epoch, with EVPAs in the upper region of the map being perpendicular to EVPAs in the lower part of the images. Apart from that, when considering this effect and applying a $90^\circ$ correction to this pronounced region (see the right part of \autoref{fig:evpa_map}), the reconstructed magnetic field direction and the RM maps turned out to be consistent between different frequencies and epochs. Additionally, a large region in the center of the 8.2 GHz EVPA map showed no signal above the 3$\sigma_P$ level. This is likely caused by changing opacity and a potentially strong increase in absorption at this specific frequency. 

\begin{table}
	\centering
	\caption{Degree of linear polarization, $m$, obtained in the center of the J component.}
	\label{tab:lin_degree_table}
	\begin{tabular}{clc} 
		\hline\hline
		Epoch & Frequency & $m$ \\ 
		& (GHz) & ($\%$)  \\
		\hline
		2005-11-26 & 22.2 & 14.2  \\   
                   & 15.4 & 5.4  \\
                   & 8.5 & 0.5   \\
                   & 8.2 & 0.5   \\
		\hline
		2007-02-09 & 22.2 & 9.5 \\   
                   & 15.4 & 8.3  \\
                   & 8.5 & 0.3  \\
                   & 8.2 & 0.7   \\
        \hline
        2007-05-23 & 22.2 & 8.6   \\   
                   & 15.4 & 7.9  \\
                   & 8.5 & 0.3   \\
                   & 8.2 & 0.5   \\
        \hline
        2007-08-03 & 15.4 & 8.6  \\
                   & 8.5 & 0.5  \\
                   & 8.2 & 0.2   \\
        \hline
	\end{tabular}
\end{table}

\begin{figure*}
	\includegraphics[width=0.42\linewidth,trim={0cm 0cm 0cm 0cm},clip]{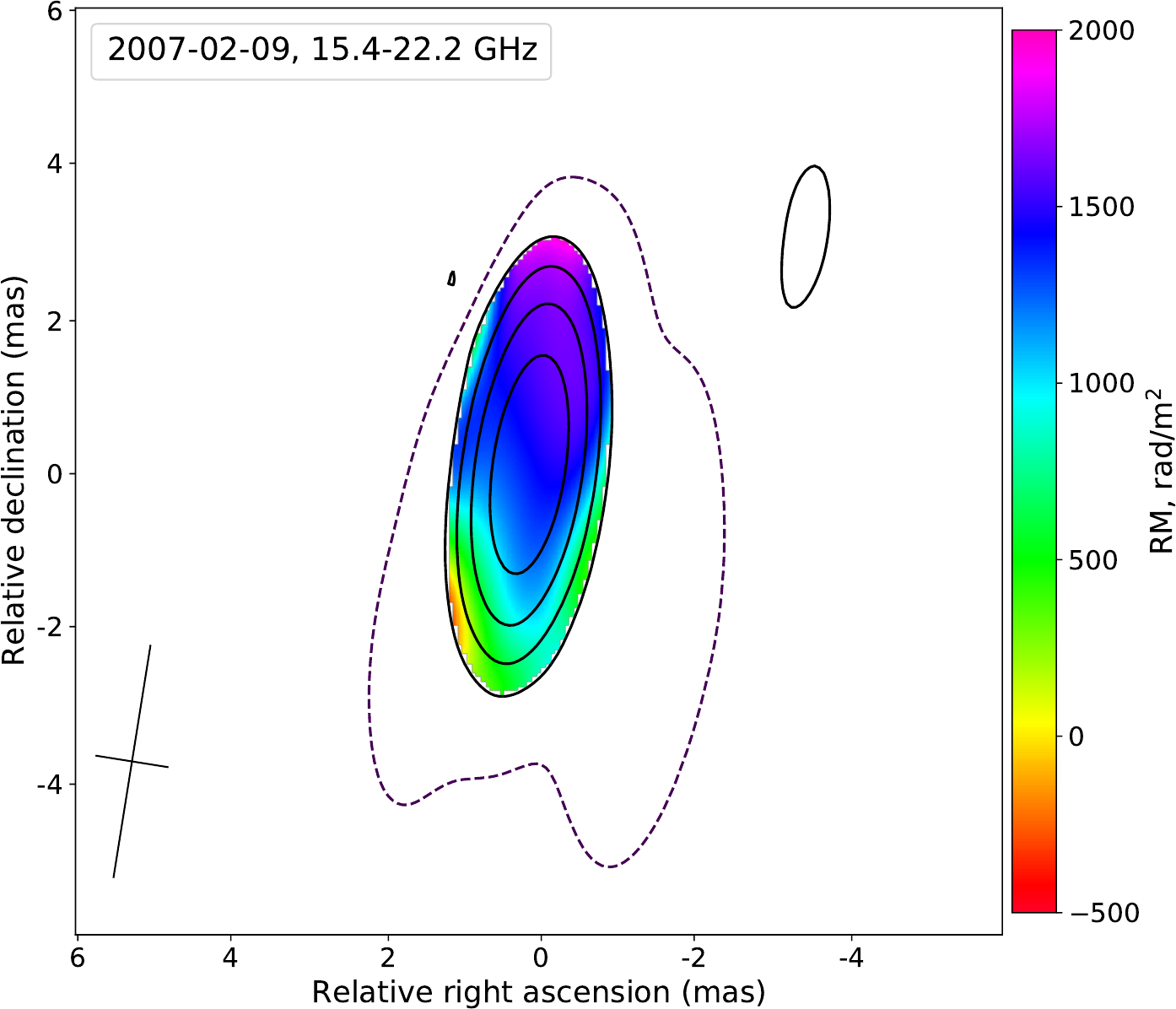}
	\hspace{1cm}
	\vspace{0.05cm}
	\includegraphics[width=0.3593\linewidth,trim={0cm 0cm 0cm 0cm},clip]{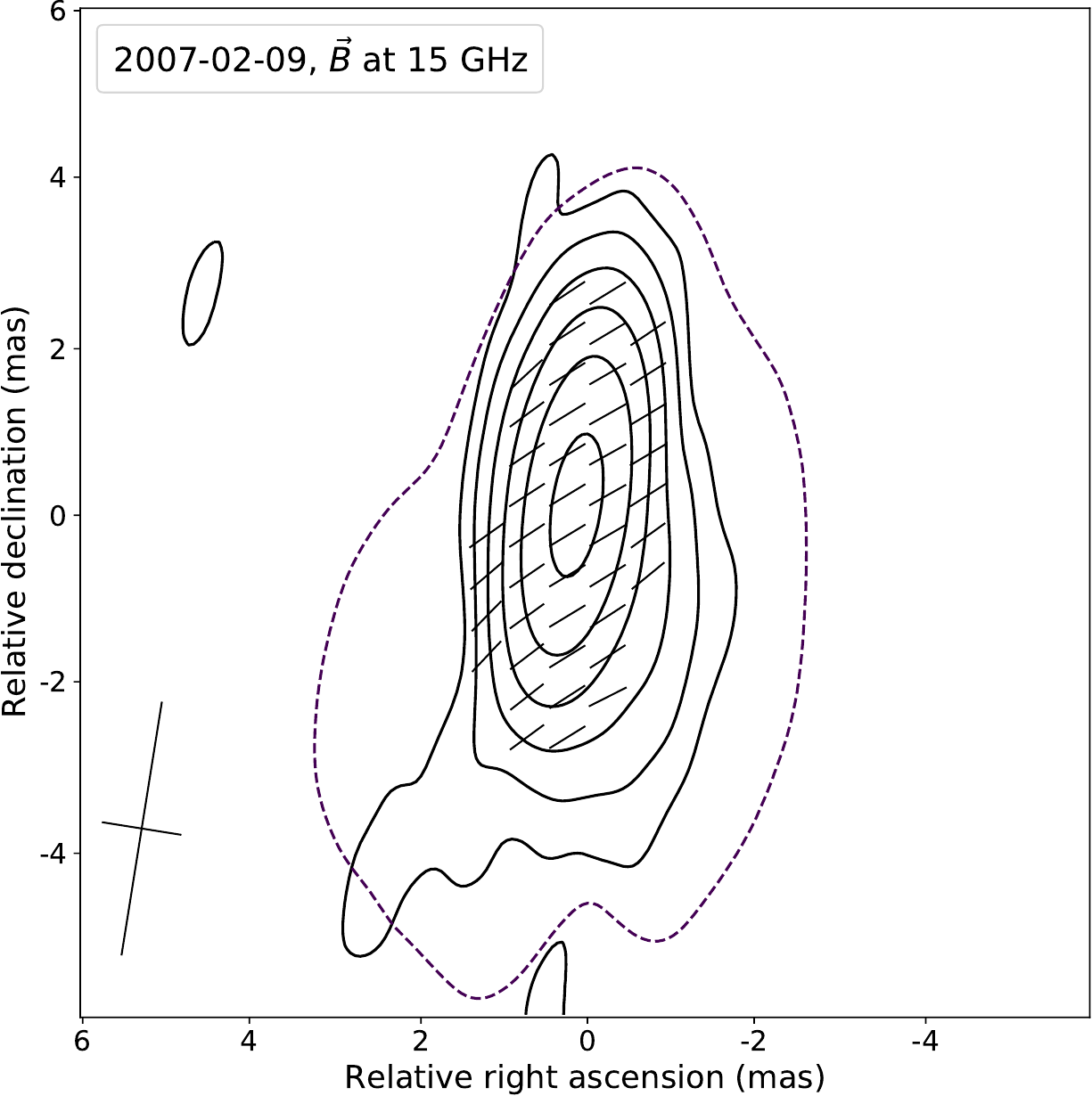}
	\includegraphics[width=0.42\linewidth,trim={0cm 0cm 0cm 0cm},clip]{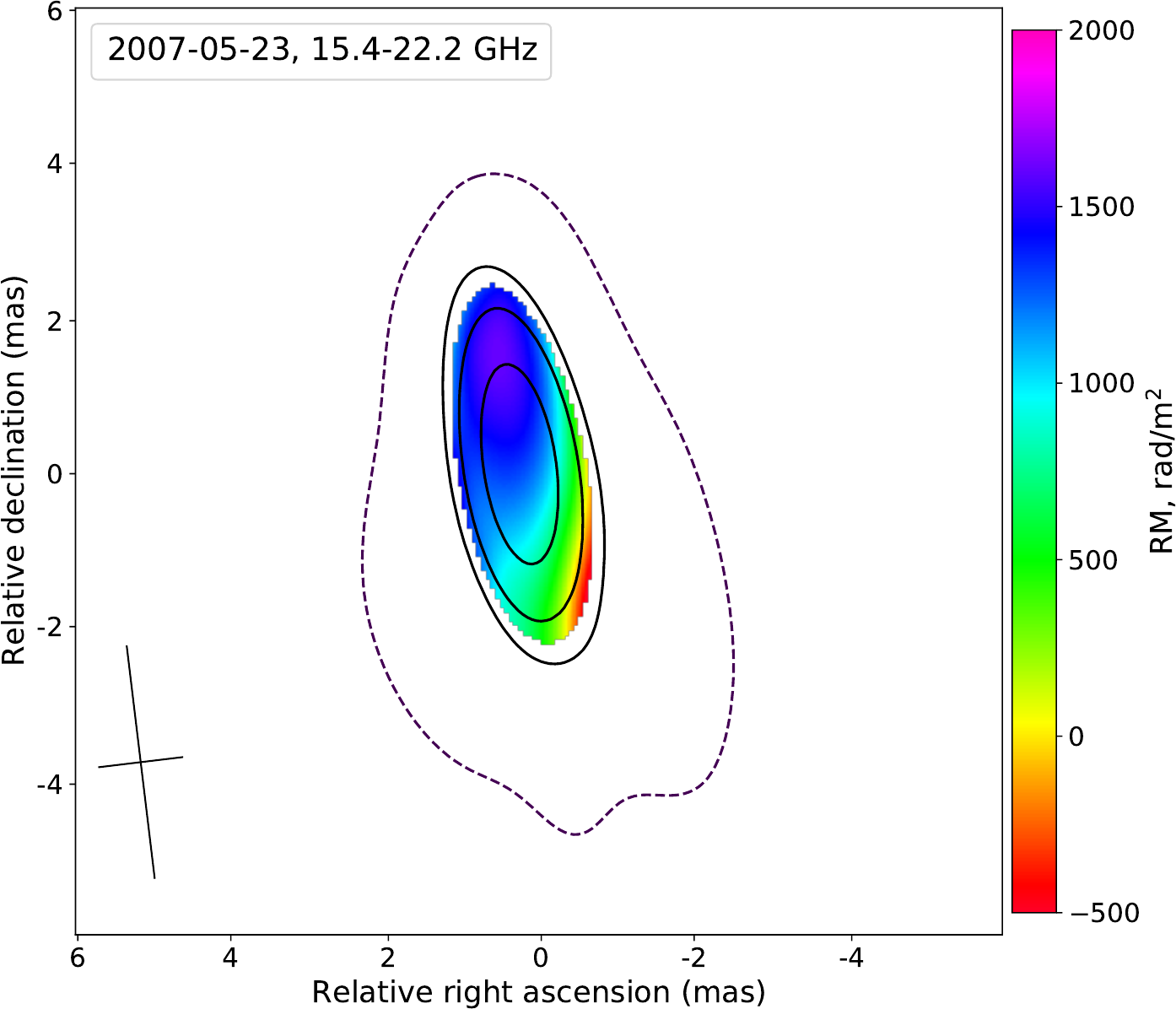}
	\hspace{1cm}
	\vspace{0.05cm}
	\includegraphics[width=0.3593\linewidth,trim={0cm 0cm 0cm 0cm},clip]{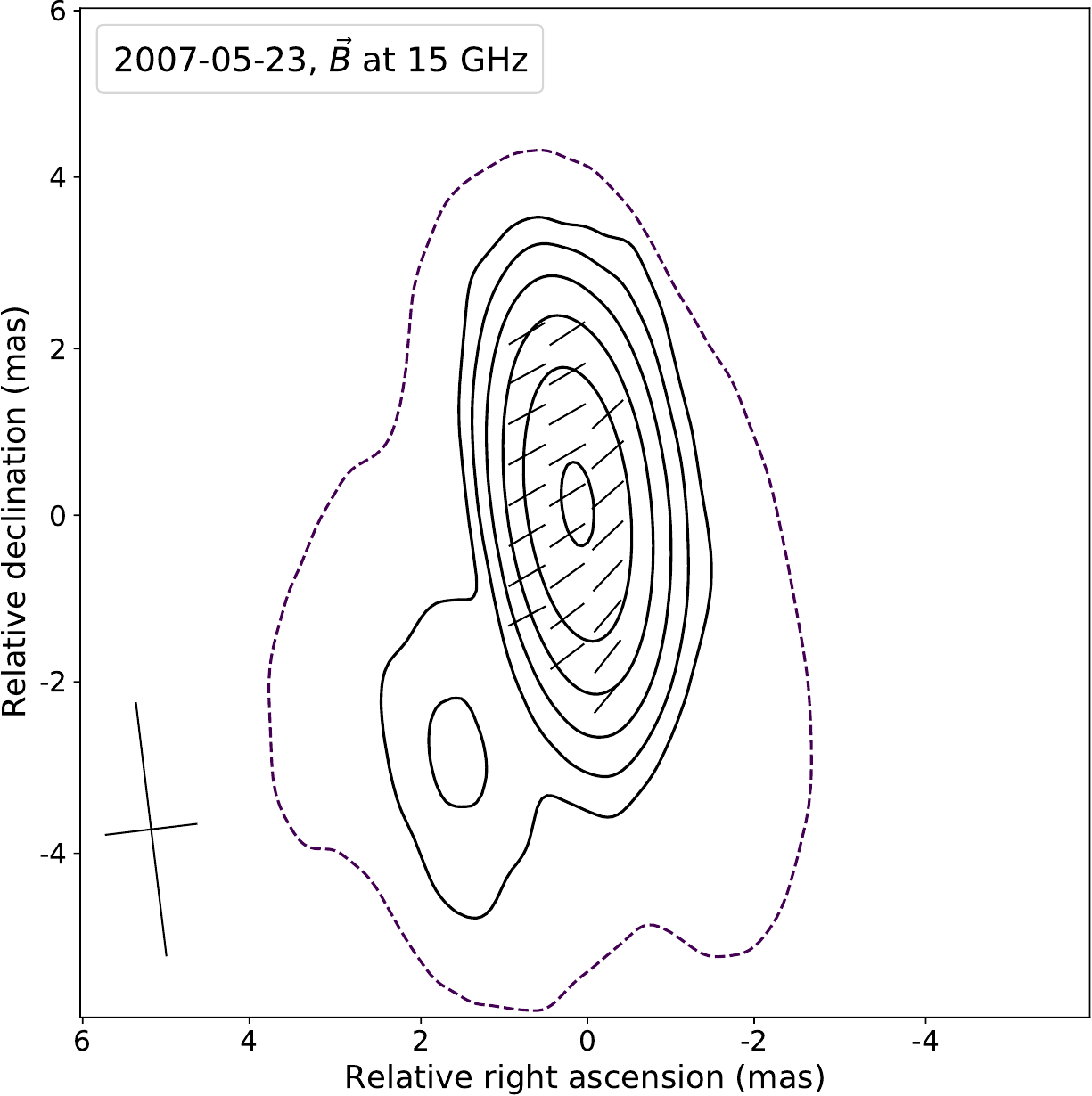}
	\includegraphics[width=0.42\linewidth,trim={0cm 0cm 0cm 0cm},clip]{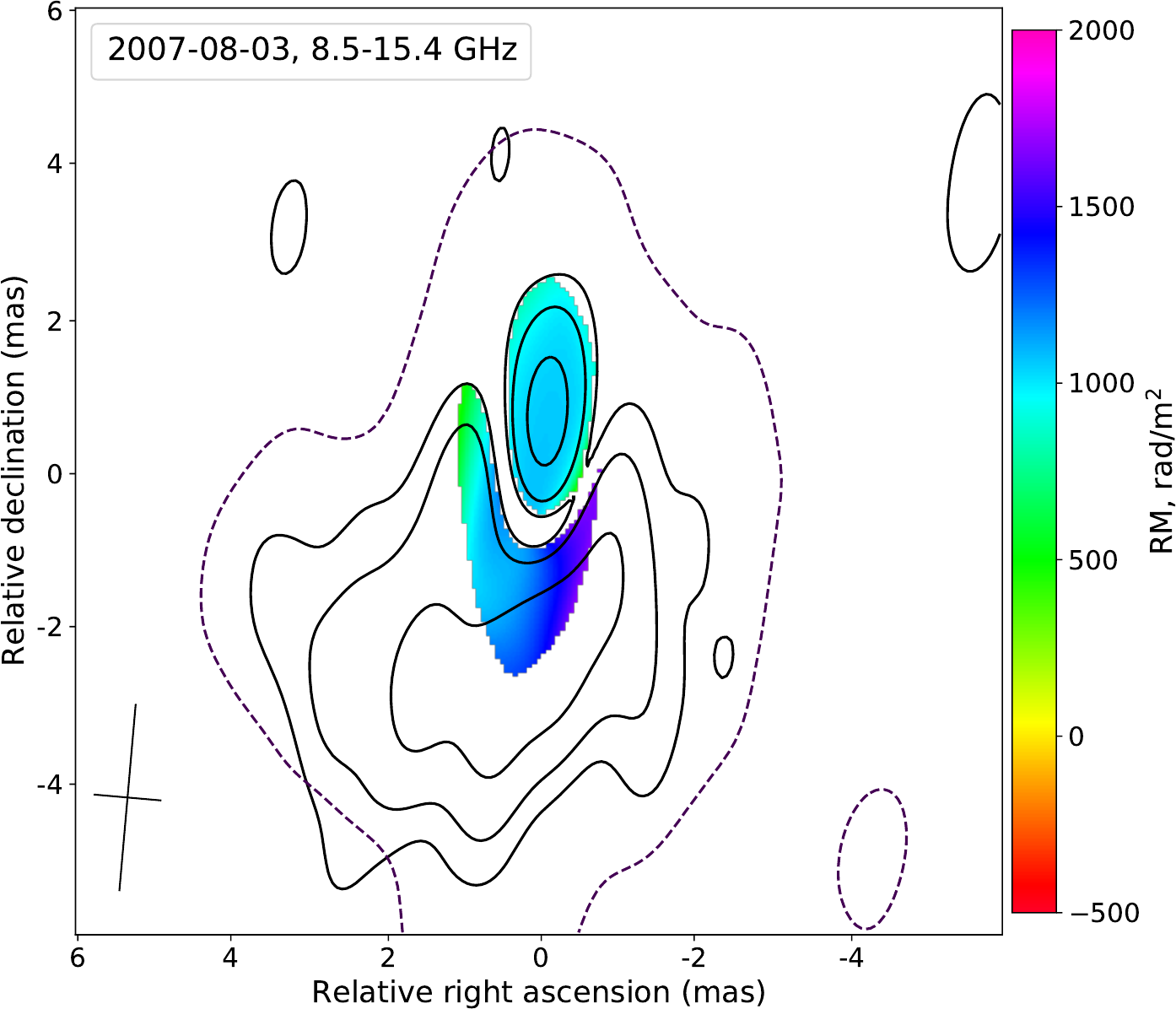}
	\hspace{1cm}
	\vspace{0.05cm}
	\includegraphics[width=0.3593\linewidth,trim={0cm 0cm 0cm 0cm},clip]{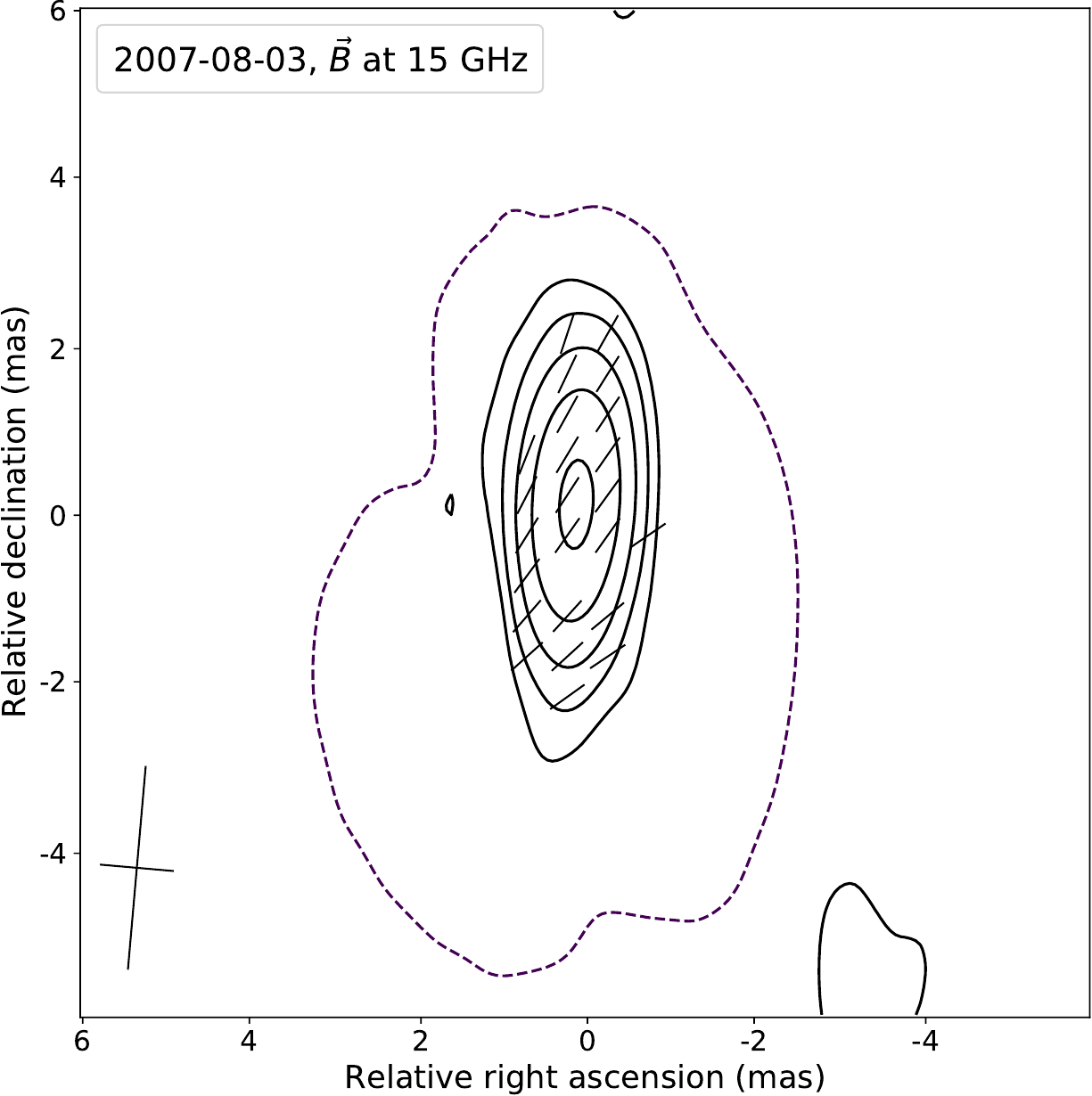}
    \caption{Faraday RM maps for the highest used frequency intervals from different epochs (left) and the magnetic field direction reconstructed at 15.4 GHz (right). The RM maps show the full intensity contour (dotted line) at 5$\sigma_I$ taken from the Stokes $I$ map at 22.2 GHz for the 15.4$-$22.2~GHz interval and at 8.5 GHz for the 8.5$-$15.4~GHz interval. The magnetic field direction maps show the full intensity contour (dotted line) at 5$\sigma_I$ taken from the Stokes $I$ map at 15.4 GHz. The RM values and the magnetic field direction are shown within the linear polarization contours starting at the 3$\sigma_P$ level with the $\times2$ steps. The restoring beam is shown as a cross in the bottom left corner at the FWHM level. The beam parameters for 2007-02-09: $B_\mathrm{maj} = 3.0$ mas, $B_\mathrm{min} = 0.9$ mas, $B_\mathrm{PA} = -9.1$ degrees; for 2007-05-23: $B_\mathrm{maj} = 3.0$ mas, $B_\mathrm{min} = 1.1$ mas, $B_\mathrm{PA} = 6.9$ degrees; for 2007-08-03: $B_\mathrm{maj} = 2.4$ mas, $B_\mathrm{min} = 0.9$ mas, $B_\mathrm{PA} = -5.0$ degrees. Parameters for the RM maps: 2007-02-09: Peak intensity value $I$ = 399 mJy/beam, $P$ peak is $39.8$ mJy/beam, $\sigma_{I}$ = 0.8 mJy/beam, $\sigma_{P}$ = 0.9 mJy/beam; 2007-05-23: Peak intensity value $I$ = 376 mJy/beam, $P$ peak is $37.4$ mJy/beam, $\sigma_{I}$ = 0.9 mJy/beam, $\sigma_{P}$ = 1.9 mJy/beam; 2007-08-03: Peak intensity value $I$ = 589 mJy/beam, $P$ peak is $5.1$ mJy/beam, $\sigma_{I}$ = 0.3 mJy/beam, $\sigma_{P}$ = 0.2 mJy/beam. For the magnetic field direction maps: 2007-02-09: Peak intensity value $I$ = 446 mJy/beam, $P$ peak is $41.6$ mJy/beam, $\sigma_{I}$ = 0.4 mJy/beam, $\sigma_{P}$ = 0.4 mJy/beam; 2007-05-23: Peak intensity value $I$ = 532 mJy/beam, $P$ peak is $47.9$ mJy/beam, $\sigma_{I}$ = 0.4 mJy/beam, $\sigma_{P}$ = 0.5 mJy/beam; 2007-08-03: Peak intensity value $I$ = 467 mJy/beam, $P$ peak is $46.7$ mJy/beam, $\sigma_{I}$ = 0.4 mJy/beam, $\sigma_{P}$ = 0.5 mJy/beam.}
    \label{fig:pol_maps}
\end{figure*}

Furthermore, we reconstructed the RM maps for frequency ranges 1.4$-$2.4~GHz, 2.3$-$5.0~GHz, 8.2$-$15.4~GHz and 15.4$-$22.2~GHz for two epochs 2007-02-09 and 2007-05-23. For the 2007-08-03 epoch, instead of taking the 8.2$-$15.4~GHz and 15.4$-$22.2~GHz ranges, we used the range 8.5$-$15.4~GHz since the 22.2~GHz data were not optimal for analysis (\autoref{tab:obs}), and polarization signal in the 8.2 GHz EVPA map is small. The RM values were calculated in the same manner as in the analysis of the 2005-11-26 epoch. We then derived the magnetic field direction images, considering the radiation regime and applying a $90^\circ$ correction when necessary. In \autoref{fig:pol_maps}, we show the rotation measure maps and the magnetic field directions for the highest frequency intervals from epochs in 2007. Apart from that, in \autoref{tab:pol_table}, we present the corresponding RM and magnetic field direction values. The results at other lower frequencies turned out to be the same within the error limits. However, we do not show them since the uncertainties in the direction of the EVPAs at low frequencies were drastically larger. In addition, the results are in good agreement between different epochs in 2007. 

We used several criteria to analyze the RM gradients in the reconstructed maps. They are described by \citet{2012AJ....144..105H} and include the jet width, which should be preferably more than two beams wide based on the simulations performed to investigate how
large spurious gradients can arise due to noise
and finite size of the restoring beam. In our study, the size of the jet width is approximately two beams, indicating we have enough resolution. Apart from that, gradients should exceed the 3$\sigma$ level, where $\sigma$ is the largest RM error. Similar criteria can be found in, e.g. \citet{2010ApJ...722L.183T} or \citet{2017MNRAS.467...83K}. For our data, the gradients turned out to be insignificant for all epochs and frequency ranges.

\begin{table}
	\centering
	\caption{Faraday rotation measure and direction of the magnetic field as projected on the plane of the sky in the dominant jet feature J.}
	\label{tab:pol_table}
	\begin{threeparttable}
	\begin{tabular}{ccccc} 
		\hline\hline
		Epoch & Frequency & RM   & Magnetic field  \\
		       & (GHz)     & ($\times 10^3$ rad/m$^2$) &   direction (deg) \\
		\hline
          2007-02-09 & 15.4 & $1.43 \pm 0.09$ &  $120 \pm 2$  \\  
                     & 22.2 &                 &  $120 \pm 1$ \\
        
        \hline
          2007-05-23 & 15.4 & $1.14 \pm 0.19$ &  $126 \pm 4$  \\  
                     & 22.2 &                 &  $126 \pm 2$ \\
        
        \hline
          2007-08-03 & 8.5 & $1.07 \pm 0.03$ &  $144 \pm 2$  \\  
                     & 15.4 &                &  $144 \pm 1$ \\
        
        \hline
        
	\end{tabular}
	\begin{tablenotes}
            \item \emph{Note.} RM and magnetic field direction were averaged over 9 pixels at the location of the map containing the brightest pixel.
            \end{tablenotes}
    \end{threeparttable}
\end{table}

\section{Discussion}
\label{sec:discus}

\subsection{Kinematics}

The structure of the quasar revealed no significant changes over a period of more than ten years. The brightest jet feature was observed during all epochs from 2005 to 2018. \autoref{fig:sp_ind} shows that the spectral index increases in this region, which must correspond to an energy gain in the emitting particles at the observed interval. The energy which goes into particle acceleration is probably being generated at the expense of kinetic energy and, taking into account that the feature is steady, we could interpret it as a strong interaction between the jet flow and a dense ISM region. The impact from a cloud could trigger a shock, which would fit in the picture of the apparent observed bend in this region where the J component is located. In addition, we observe the component S between the core C and the brightest jet feature, J$_1$ in 2005 and J in 2007, at the highest frequencies (see \autoref{fig:components}). The low resolution during other epochs did not allow us to detect it in 2017 and 2018. This component was located at a distance of $\sim$0.7 mas from the core in 2005 and May and August of 2007. According to its properties, we suggest that this component could be a stationary recollimation shock wave. Its position shifted a little further downstream in February of 2007 with the distance from the core becoming $\sim$1 mas. Such behaviour could be a consequence of the standing shock being dragged downstream and then returning to its initial position upstream \citep[e.g.][]{1997ApJ...482L..33G,2016A&A...588A.101F}. However, changes in the opacity of the core region could also affect the relative positions of the components.

The two components J$_0$ and J$_1$ observed in 2005 (see ~\autoref{fig:components}) could hint that there were two shock waves in 2005, one of which was stationary and the other was travelling. In this scenario, J$_1$ in 2005 could be cross-identified with the J component in 2007 by the same distance to the core – assuming that the opacity effects did not affect the core positions. This would suggest that J$_0$ was the travelling component. However, it is not clear why the J$_1$ flux density in 2005 was not consistent with the flux densities of J in 2007 (see \autoref{tab:gauss_comp}). One feasible explanation could be related to nonlinear effects of interacting shock waves, which is the scenario we favour (see also further discussion of the spectral peak values evolution). 

If the J and J$_0$ components were the same ones, the apparent velocity would be $(8.9 \pm 1.9)c$, where $c$ is the speed of light, which again made use of an assumption that the core position was stationary. In principle, the core position can change due to significant nuclear flares \citep[e.g.][]{2019MNRAS.485.1822P}. The timescale of such changes may align with our observed timescales, on the order of a year. However, considering the relatively small variation in flux density in our core component, we propose that the assumption of a stationary core position in this context is justified.

Although superluminal motion is often found in the jets \citep[e.g.][]{2003ASPC..299..117K,2016AJ....152...12L}, in this case, the distance between the core and the component J would have decreased, implying a backward motion of the brightest jet feature. Such behavior would be exceptionally peculiar at these velocities, as slower motions are typically observed in sources exhibiting inward movements \citep[e.g.][]{2013AJ....146..120L, 2016AJ....152...12L, 2021ApJ...923...30L}. At the same time, it would not be consistent with the results obtained in \autoref{sec:kinem}. Nevertheless, significant superluminal component motions towards the core can sometimes occur \citep[e.g.][]{2015PKAS...30..429S}. They are mostly explained by virtue of highly curved jets geometry or emerging unresolved components in the core. In the context of a highly curved geometry, Sawada-Satoh~et~al.\ suggest that if the apparent inward motion is due to a highly curved trajectory, the flux density of the corresponding moving component would exhibit time variability due to a variation in the geometrical-origin Doppler factor. However, the radio light curve in their study reveals a constant flux density during the inward motion, rendering this scenario less likely. In our observation, significant changes in flux density occur when J is associated with J0. Simultaneously, in this situation, the jet should curve back towards the line of sight \citep[e.g.][]{2003ASPC..299..117K}, making emission in the southeast direction unlikely.

Apart from that, clear helical patterns are sometimes observed in AGN jets \citep[e.g.][]{1999NewAR..43..699H, 2004A&A...417..887H, 2011A&A...529A.113Z, 2012ApJ...749...55P, 2021A&A...654A..27B,2022ApJ...924..122G,2023NatAs.tmp....8F}{}{} where a curved jet geometry is noticeable and could potentially explain the inward motion in our scenario. Typically, these instances exhibit drastic changes in the position angle of jet components \citep[e.g.][]{2021ApJ...923...30L}. Contrarily, in our case, the position angle remains relatively stable over extended periods, leading us to favour alternative scenarios. Moreover, the field is usually perpendicular to the direction of propagation, whereas in our case it is aligned with the hypothetical interaction region. These observations suggest that this scenario may not fully apply for our data.

Finally, the last option is that the two components in 2005 could not represent any real physical features, as discussed in \autoref{sec:tot_int_struc}. The numerical simulations of the interaction between a stationary shock and a traveling one \citep[e.g.][]{2016A&A...588A.101F}  and further high-frequency VLBI observations could potentially help us distinguish between different scenarios. 

\subsection{Spectral Changes and Flux Density Variability in the Core and Jet Feature} 

The observed changes in the flux density in the core and the jet feature suggested that occasionally new material was ejected in the jet. Apart from the visible effects on the VLBI variability images, this could also be a natural explanation for the variation of flux densities in the integrated spectrum (see \autoref{fig:light_curve}).

In principle, scattering effects could impact the flux variability of the quasar. As elucidated in \citetalias[][]{2022MNRAS.510.1480K}, an analysis of the size-frequency dependence for the quasar was conducted, primarily focusing on the core at high frequencies (15 GHz and 22 GHz). The findings revealed a good agreement with the conical synchrotron jet model, where core size depends on frequency as $\propto \nu^{-1}$ \citep[][]{1979ApJ...232...34B}, suggesting that scattering effects were not dominant at these frequencies. In this paper, upon extending the analysis to other epochs, a deviation from the conical synchrotron jet model was observed, hinting at the potential presence and influence of scattering. While extreme scattering events can occasionally modify the flux by a factor of a few, exhibiting a distinct U-shaped pattern \citep[e.g.][]{2018ASPC..517..491V}, they generally do not contribute significantly to the overall observed variability. Furthermore, beyond these extreme events, scattering does not notably alter the flux, as per typical scattering behaviour. Apart from these considerations, strong scattering is deemed not very likely given the sky coordinates of the quasar \citep[e.g.][]{2015MNRAS.452.4274P,2022MNRAS.515.1736K}, leading us to maintain the assumption that scattering does not significantly impact our main results. 

The ($\nu_\mathrm{m}$ $-$ $S_\mathrm{m}$)-plane shows two distinct stages (see \autoref{fig:peak_fl_fr}). The first stage, lasting from 1997 to $\sim$2010, displays a general trend of both the flux density and the peak frequency dropping, indicative of an adiabatic stage in the classical shock-in-jet model \citep[][]{1985ApJ...298..114M}. The subsequent stage presents a scenario where the flux density remains roughly constant, but the peak frequency increases. This latter behavior is not typical for the classical model, given the recognized loss mechanisms such as adiabatic, synchrotron, or inverse Compton losses. Such observations suggest a more intricate picture, hinting that the simple shock-in-jet model may not fully encapsulate the dynamics at play in the jet. Moreover, the high errors of $\nu_\mathrm{m}$ and $S_\mathrm{m}$ do not allow us to distinguish between other models of shock-shock interactions and extract any parameters. In \cite{2016A&A...588A.101F}, the authors show that the evolution of the ($\nu_\mathrm{m}$ $-$ $S_\mathrm{m}$)-plane depends on whether the spectral peak is produced at the shock-perturbed region in the case of a conical expanding jet or in the case of shock-shock interactions. According to these results, the evolution in the aforementioned plane is clockwise in the latter case and counter-clockwise in the former (see Fig.~11 in that paper). Although the numerical simulations simplify the scenarios, and the spectra that we observe correspond to the whole source rather than the J components, we can derive interesting conclusions from a direct comparison. Indeed, the global evolution in the ($\nu_\mathrm{m}$ $-$ $S_\mathrm{m}$)-plane (see \autoref{fig:peak_fl_fr}) is in the counter-clockwise direction, which means that the spectral peak is probably produced by the evolution of a perturbation along the conically expanding region. However, in 2004, the peak coordinates displace to the right in the plane (see points in black squares in \autoref{fig:peak_fl_fr}) and come back to follow the original track along the C-shaped evolution with oscillations to lower peak frequencies. The sudden rightward motion is perhaps due to the interaction between the components J0 and J1. It is possible that precisely in 2005, we started to see these shocks separating. The rest of the time-span between 1997 and 2020, the evolution of the spectral peak would be dominated by the expansion of the travelling perturbation along the jet, until the component fades out. At this point, the spectral peak could be produced either by the core or by the bright stationary J~component.

\subsection{Rotation Measure Results and Implications}

We note that the RM values dropped significantly from above 6000 rad/m$^2$ in 2005 to around 1000 rad/m$^2$ in 2007. This behaviour is not common for jets but possible for various reasons, and it has been observed before \citep[e.g.][]{2001ApJ...550L.147Z}. One possibility to explain this could be that properties of the external medium in the screen changed significantly, i.e.\ its density, magnetic field, or the direction of the magnetic field, causing a change in the projected value along the line of sight.
In this case, the hypothetical colliding external cloud should be rather small to cause such changes over year-long periods in the observed RM. When we tried to calculate its size, taking into account causality, it turned out to be around 0.3 pc. If projection effects were added, the cloud size would have parsec scales. Nevertheless, both the magnetic field and the density of the interacting regions might have dropped within the span of just over one year separating the observing epochs, causing this change. Given the causality argument, this implies that the region undergoing these changes must be smaller than just over one light-year in size. It is feasible that the hypothetical interaction between the jet and a dense cloud affected the cloud properties and the electron density decreased. The numerical simulations by \cite{2012A&A...539A..69B} showed that the interaction between a jet and a cloud triggers a shock in the latter, heating it and causing its expansion. The cloud thus becomes less dense. Apart from that, under this interpretation, we can predict that the RM values should keep falling and, at some point, the component J will fade out.
Within this scenario, quasar 0858$-$279 represents a good candidate to show shock-ionization lines as those observed in other young sources \citep[e.g.][]{2005A&A...436..493L,2006A&A...447..481L,2009MNRAS.400..589H,2016MNRAS.455.2453M,2021A&A...656A..55M}.

Alternatively, a different part of the jet might have been illuminated, leading us to measure the RM values in that distinct region. However, the distance between the most illuminated parts of the jet should not exceed the distance between the J$_0$ and J$_1$ components, which is $\sim$0.2 mas (or 1.5 pc), quite short for such drastic changes. If we assume that there is one cloud with constant properties and a size of more than 1 pc, this option is highly unlikely.

The direction of the magnetic field is also considerably altered: it became perpendicular to the propagation direction of the jet from the central engine to the observed bend (\autoref{fig:pol_maps}). This could fit well into the picture of two interacting shock waves, in the sense that the polarized emission could be concentrated at the head of the moving shock.  However, the interaction between the jet and the Faraday screen could also be the reason for a change in polarization properties.

\section{Summary}
\label{sec:summary}

The GPS quasar 0858$-$279 initially exhibited several bizarre peculiarities, including strong variability, which could not be explained by its seemingly stable structure in the early VLBI images. In the analysis of the 2005-11-26 epoch, we were able to resolve a typical core-jet structure. We also came to the conclusion that a shock wave is formed in the jet via a strong interaction with the ambient medium. That made the emission from this part brighter than from other parts of the source. 

In this paper, we have presented multi-epoch analysis of the quasar with additional six epochs of observations. The brightest jet feature was detected during all epochs covering a range of almost 15 years. VLBI-\textit{Gaia} astrometry confirmed the interpretation of the source structure and the identification of the core. The variability of emission demonstrated that most of the changes occurred in the brightest jet feature, yet for some epochs, flux densities varied slightly in the core region. Apart from that, the spectral properties of the source changed within a year-long period. This included flattening of the spectrum in the jet feature -- components~J -- and spectrum becoming more inverted in the core region. Furthermore, we studied the polarization properties for three epochs in 2007. The most noticeable changes were the RM values decreasing from 6000 rad/m$^2$ to 1000 rad/m$^2$ in one year.

We studied the spectral evolution of the source along the epochs and found that it fits well into the picture of evolving perturbation along the conically expanding jet, presumably between the components C and J. However, we observe rapid changes in peak frequency and flux density around 2005. This is interpreted as a hint of interaction between the evolving perturbation and the bright component J, i.e. a possible shock-shock collision, as supported by numerical simulations \citep{2016A&A...588A.101F}. 

The changes in the RM values can be interpreted as being produced by the changes in the screening medium. In particular, this could be induced in the hypothetical ISM cloud by the shock. In this frame, as the cloud is shocked and heated, it expands and becomes more dilute, changing the electron densities and therefore, the measured Faraday rotation. Interestingly, this is also in good agreement with the interpretation of a strong collision between the jet and a dense-cloud, which would act as a Faraday screen. In this scenario, we expect the RM to keep falling in time and, at some point, the component J to become dimmer. 

In one year period, we also observe a change in the direction of the jet magnetic field, becoming perpendicular to the propagation of the jet before its observed bend. If we take into account that the jet kinematics and spectral evolution could also be interpreted in terms of travelling perturbation interacting with recollimation, standing shock at the location of the component J, the changes in the magnetic field direction could be associated to the reinforcement of the perpendicular components by the travelling shock, after the passage.  

Further VLBI observations with improved uv-coverage with polarization analysis, search for hints of shock ionization of the ISM, as indicative of jet-medium interaction, and numerical simulations could shed light on the evolution of this peculiar quasar.

\section*{Acknowledgements}

We thank Nicholas MacDonald for reading the manuscript and useful comments. We thank E. Bazanova for the English language editing of this manuscript.

YYK was supported by the M2FINDERS project which has received funding from the European Research Council (ERC) under the European Union’s Horizon 2020 Research and Innovation Programme (grant agreement No~101018682).
MP acknowledges support by the Spanish Ministry of Science through Grants PID2019-105510GB-C31, PID2019-107427GB-C33 and from the Generalitat Valenciana through grant PROMETEU/2019/071. 

The VLBA is an instrument of the National Radio Astronomy Observatory, a facility of the National Science Foundation operated under the cooperative agreement by Associated Universities, Inc. 
This work made use of the Swinburne University of Technology software correlator, developed as part of the Australian Major National Research Facilities Programme and operated under license \citep{2011PASP..123..275D}.
The observations with the SAO RAS telescopes (RATAN-600) are supported by the Ministry of Science and Higher Education of the Russian Federation. 
This research made use of NASA’s Astrophysics Data System Bibliographic Services.
This research made use of the data from the MOJAVE database, which is maintained by the MOJAVE team \citep{2018ApJS..234...12L} and the University of Michigan Radio Astronomy Observatory. 
This work has made use of the data from the European Space Agency (ESA) mission \textit{Gaia} (\url{https://www.cosmos.esa.int/gaia}), processed by the \textit{Gaia} Data Processing and Analysis Consortium (DPAC, \url{https://www.cosmos.esa.int/web/gaia/dpac/consortium}). Funding for the DPAC has been provided by national institutions, in particular the institutions participating in the \textit{Gaia} Multilateral Agreement.

\section*{Data Availability}

The correlated data underlying this article are available from the public NRAO archive\footnote{\url{https://data.nrao.edu/}}, project codes are BK128 and BK133.
The fully calibrated visibility and the image FITS files for the target source are made available online by us in CDS\footnote{\url{https://cds.u-strasbg.fr/}}. We made use of the publicly available Astrogeo VLBI FITS image database\footnote{\url{http://astrogeo.smce.nasa.gov/vlbi_images/}}.



\bibliographystyle{mnras}
\bibliography{0858}

\bsp	
\label{lastpage}
\end{document}